\documentclass[a4paper,11pt]{article}

\usepackage{jheppub}

\usepackage[T1]{fontenc} 
\usepackage{dsfont}
\usepackage{framed}

\usepackage[makeroom]{cancel}

\allowdisplaybreaks[1]

\newcommand{\nn}{\nonumber\\}
\usepackage[T1]{fontenc}
\usepackage{caption}
\usepackage{subcaption}
\usepackage{dsfont}
\usepackage{extarrows}
\usepackage[aligntableaux=center,boxsize=0.8em]{ytableau}

\allowdisplaybreaks[1]

\def\({\left(}
\def\){\right)}
\def\[{\left[}
\def\]{\right]}
\def\<{\left\langle}
\def\>{\right\rangle}

\def\zb{\bar{z}}

\def\a{\alpha}
\def\b{\beta}
\def\g{\gamma}
\def\G{\Gamma}
\def\d{\delta}
\def\D{\Delta}
\def\e{\epsilon}
\def\z{\zeta}

\def\l{\lambda}

\def\m{\mu}

\def\p{\pi}

\def\t{\tau}

\def\nn{\nonumber\\}
\def\pa{\partial}

\newcommand*{\Scale}[2][4]{\scalebox{#1}{$#2$}}

\begin{document} 
	
\title{Anomalous dimensions of partially conserved higher-spin currents from conformal field theory: bosonic $\phi^{2n}$ theories}
	
\author{Yongwei Guo}
\author{Wenliang Li}
\emailAdd{liwliang3@mail.sysu.edu.cn}
\affiliation{School of Physics, Sun Yat-sen University, Guangzhou 510275, China}
	
\abstract{
	In the free $\Box^k$ scalar conformal field theory, 
	there exist conserved and partially conserved higher-spin currents.  
	We study their anomalous dimensions associated with $\phi^{2n}$ interaction in the $\epsilon$ expansion.  
	We derive general formulas for the leading corrections from the conformal multiplet recombination,  
	and verify their consistency with crossing symmetry using the Lorentzian inversion formula.
	The results are further extended to the O($N$) models.
}
	
\maketitle

\section{Introduction}
\label{sec:Introduction}
A natural generalization of the free scalar conformal field theory (CFT) is 
to consider higher powers of the Laplacian. The higher-derivative action reads
\begin{align}
S\propto\int\mathrm{d}^{d}x\;\phi\,\Box^{k}\phi\,.
\label{free-action}
\end{align} 
The properties of the free $\Box^k$ scalar CFT have been studied in some detail in 
\cite{Brust:2016gjy}. (See also \cite{David:1992vv,Osborn:2016bev,Gracey:2017erc}.)
For $k=1$, this is the standard free scalar CFT. 
For $k>1$, we have a nonunitary CFT as $\D_\phi=\frac d 2-k$ violates the unitarity bound.\footnote{ The unitarity bound is derived in the standard quantization of CFT.
There may exist certain quantizations where the $k>1$ theories are unitary, but the energies could be unbounded from below. }
Nevertheless, the higher-derivative CFT exhibits some interesting features,  
such as generalized conservation laws and extended higher-spin symmetry. 
Besides the well-known higher-spin conserved currents, 
there exist $(k-1)$ towers of partially conserved currents. 
They are also called multiply conserved currents in \cite{Brust:2016gjy}. 
Curiously, these additional towers of higher-spin operators do not vanish by contracting with one derivative. 
Instead, they vanish upon the action of multiple derivatives. 
Schematically, the generalized conservation laws read
\begin{align}
\pa J\neq 0\,,\quad
\pa\ldots\pa J=0\,,
\end{align} 
where indices are suppressed for simplicity. 
More details will be provided in Sec. \ref{sec:Partially conserved current}. 

It is straightforward to introduce the O($N$) global symmetry\footnote{Here we assume the scalar fields are real. For complex scalars, we have U($N$) global symmetry.} 
by considering $N$ copies of the free action \eqref{free-action}. 
In the AdS/CFT correspondence \cite{Maldacena:1997re,Gubser:1998bc,Witten:1998qj}, 
the large-$N$ (higher-derivative) CFT is expected to be dual to a (partially) massless higher-spin gravity on AdS$_{d+1}$ \cite{Klebanov:2002ja,Bekaert:2013zya}.
The latter is also known as the type-A higher-spin theory (with higher depth) \cite{Vasiliev:1990en,Vasiliev:1992av,Vasiliev:2003ev,Deser:1983mm,Higuchi:1986wu,Deser:2001us,Joung:2012rv,Joung:2012hz,Alkalaev:2014nsa,Brust:2016zns}.\footnote{In accordance with the nonunitary nature of the $k>1$ CFTs, 
the partially massless gravities are also nonunitary. 
See \cite{Joung:2014aba} for the case of interacting spin-$2$ fields.}
A more precise dictionary is that 
the single trace (partially) conserved currents of the boundary CFT 
are dual to (partially) massless higher-spin particles in the bulk \cite{Dolan:2001ih}.  

In this work, we will consider the $\phi^{2n}$ deformation of the free $\Box^k$ CFT.\footnote{ In this work, we only consider the perturbative $\e$ expansion.
To the best of our knowledge, it is an open question whether the $k>1$ theories make sense nonperturbatively. }
For $k=1$, the $\phi^{2n}$ interactions with $n=2,3,4,\dots$ are generalizations of the $\phi^4$ Wilson-Fisher fixed point with $n-1$ relevant singlet scalar operators.
For instance, they describe the behavior of critical ($n=2$), tricritical ($n=3$), tetracritical ($n=4$) phenomena. 
We are interested in the higher-derivative generalizations of these multicritical theories. 
The upper critical dimension is given by 
\begin{align}
d_\text{u}=\frac{2nk}{n-1}.
\end{align}
Above the upper critical dimension, i.e., $d>d_\text{u}$, they are expected to be described by mean field theory. 
The $\phi^{2n}$ interaction is marginal at $d=d_\text{u}$,   
but it can induce a relevant deformation for the generalized Gaussian fixed point at $d<d_\text{u}$.
When $k$ and $n-1$ have a common divisor, 
one can also introduce deformations associated with derivative interaction terms.\footnote{
To construct $\mathbb{Z}_{2}$-even derivative terms, we replace 
an even number of $\phi$'s in $\phi^{2n}$ with an even number of derivatives. 
Note that the scaling dimension of $\phi$ is $\frac{k}{n-1}$ at $d=d_{\text{u}}$. 
For $\pa\dots\pa\phi^{2(n-j)}$,  
the scaling dimension of the derivatives, $2j\frac{k}{n-1}$, should be an even integer. 
If $k$ and $n-1$ have no common divisor, then we have $j=t(n-1)$, where $t$ is a positive integer.
But the remaining number of $\phi$ in the derivative term is $2n-2t(n-1)\leqslant2$, 
so it can only be the kinetic term.
On the other hand, if $k$ and $n-1$ have a common divisor, we can assume that $\frac{k}{n-1}=\frac{a_{1}}{a_{2}}$, where the integers $a_{1}$ and $a_{2}$ satisfy $a_{1}<k$ and $a_{2}<n-1$.
Then we can take $j=ta_{2}$, corresponding to  $\pa^{2ja_{1}}\phi^{2(n-ta_{2})}$.
In particular, when $t=1$, the number of $\phi$ satisfies $2n-2ta_{2}>2n-2(n-1)=2$, and thus this is not a kinetic term. See \cite{Safari:2017irw,Safari:2017tgs} for more discussions about derivative interactions.
}
To reach an IR fixed point, it might be inconsistent to turn on only the $\phi^{2n}$ deformation.
We will not consider these special cases and assume that $k$ and $n-1$ have no common divisor.

For small $\e=d_\text{u}-d$, the renormalization group (RG) flow induced by $\phi^{2n}$ is short between the free and interacting fixed points, 
so one can study the interacting theories in the perturbative $\e$ expansion.\footnote{ The O($N$) fixed points might be RG-unstable.
For example, in the case of $k=1$ and $n=2$, the lowest rank-4 symmetric traceless scalar made up of four fundamental fields becomes relevant when $N$ is greater than $N_c$. 
The value of $N_c$ has been studied 
earlier by the high-order $\e$ expansion \cite{Carmona:1999rm} and 
more recently by the nonperturbative numerical bootstrap \cite{Chester:2020iyt}. 
See also the references in \cite{Carmona:1999rm,Chester:2020iyt}. 
The RG instability of the $k=1$ and $n=3$ theory has also been examined in  
\cite{Osborn:2017ucf}. 
To the best of our knowledge, the RG stability of the O($N$) fixed points has not been discussed for the $k>1$ cases.
It would be interesting to study the RG stability of these higher-derivative theories. }
The traditional approach is to use diagrammatic methods to compute the corrections based on the Lagrangian formulation, 
without using the emergent conformal symmetry at the fixed points.

In light of the revival of the $d>2$ conformal bootstrap program \cite{Ferrara:1973yt,Polyakov:1974gs,Rattazzi:2008pe, ElShowk:2012ht, El-Showk:2014dwa, Kos:2014bka, Kos:2016ysd,Poland:2018epd}, 
we will study these higher-derivative multicritical theories 
directly based on the assumptions of conformal symmetry and some consistency requirements. 
Conformal symmetry implies that 
\begin{itemize}
\item
The states are organized into conformal multiplets.\footnote{
In nonunitary CFTs, there exist reducible but indecomposable representations, such as the Jordan block form \cite{Gurarie:1993xq}.  
For $\Box^k$ CFTs, one also encounters ``extended Verma modules'' in some special cases  \cite{Brust:2016gjy}.}
\item
Correlation functions take certain specific functional forms. 
\end{itemize}
Besides the symmetry constraints, we will consider two consistency requirements:
\begin{itemize}
\item
The limit $\e\rightarrow 0$ is smooth.
\item
Operator product expansion (OPE) is associative. 
\end{itemize}
The first requirement is intrinsic to the $\e$ expansion approach and 
leads to the method of conformal multiplet recombination \cite{Rychkov:2015naa}.
The second requirement gives rise to crossing constraints, 
i.e., the conformal block summations in different channels should give the same correlator. 
In fact, the first requirement is implicitly using the second one 
because the free theory itself is a consistent solution of crossing constraints. 
Below, we will elaborate on these two points. 

A CFT is characterized by the data of local operators, 
such as their scaling dimensions and OPE coefficients, i.e., $\{\D_i,\l_{ijk}\}$. 
For a free theory, including the $k>1$ generalization, the CFT data can be derived from Wick contractions. 
Knowing all the explicit numbers, 
we can think of it as one of the many consistent solutions of the CFT axioms and 
forget about the interpretation in terms of a concrete Lagrangian of free scalar. 
For example, the dynamical information of a scalar primary with $\D=3$ is completely given by the OPE coefficients involving this scalar, 
and we do not need to know if it is a composite operator of a more fundamental scalar with $\D=1$. 
For simplicity, this abstract CFT will still be called the free CFT, 
in the sense that $\{\D_i,\l_{ijk}\}$ coincide with those of the free theory, 
and we will refer to the operators by the corresponding operators in the concrete free theory representation. 

When considering a deformation of the free CFT, 
the scaling dimensions and OPE coefficients are functions of $d$ 
and they should reduce to the free CFT values in the Gaussian limit $d\rightarrow d_\text{free}$, 
i.e., when $d$ is set to the dimensions of the undeformed free CFT.  
However, an arbitrary deformation at $d=d_\text{free}-\e$ is singular in the Gaussian limit $\e\rightarrow0$.  
This is due to the existence of zero-norm states implied 
by conformal symmetry for specific scaling dimensions. 
For instance, a scalar field $\phi$ saturating the unitarity bound, i.e., $\D_\phi=d/2-1$, should obey the equation of motion $\Box \phi=0$ from purely group-theoretical arguments, 
without resorting to Lagrangians. 
If a would-be zero-norm state has finite OPE coefficients, 
then the Gaussian limit $\e\rightarrow 0$ of correlators may contain divergent contributions 
after inserting $1=\sum_{\mathcal O}\sum_{\a,\b=\mathcal O, P\mathcal O, PP\mathcal O,\dots }|\a\rangle
\langle\a|\b\rangle^{-1}\langle\b|$, where $P$ are momentum generators.  
To have a regular limit, the OPE coefficients associated with this dangerous state should also vanish. 

A subtlety arises as a change in the normalization of a would-be zero-norm state 
may lead to a finite norm as well as finite OPE coefficients in the Gaussian limit, 
so it remains a physical state and gives rise to finite contributions in the free OPEs, 
as in the standard l'H\^opital's rule. 
This is precisely the case of the Wilson-Fisher CFT with $k=1$ and $n=2$, 
whose $\{\D_i,\l_{ijk}\}$ in the Gaussian limit is identical to the free CFT. 
For example, the descendant $\e^{-1}\Box\phi_\text{WF}$ with $\D=3+O(\e)$ becomes 
a  scalar primary with $\D=3$ in the Gaussian limit
\begin{align}
\lim_ {\e\rightarrow 0}\e^{-1}\Box\phi_\text{WF}\propto\phi^3_\text{free}\,,
\end{align}
corresponding to $\phi^3$ in the free CFT. 
Therefore, the two free multiplets at $d=d_\text{u}$ recombine into one Wilson-Fisher multiplet at $d=d_\text{u}-\e$
\begin{align}
	\{\phi\}_\text{WF}\approx\{\phi\}_\text{free}+\{\phi^{3}\}_\text{free}\,,
\end{align}
which is called the conformal multiplet recombination \cite{Rychkov:2015naa}. 
To have a smooth limit $\e\rightarrow 0$, 
the deformed data should reduce to the free data, leading to nontrivial constraints on the leading corrections. 
The solution is precisely the Wilson-Fisher data.
This also generalizes to the deformation around some special scaling dimensions below the unitarity bound, 
corresponding to the $\Box^k$ free scalar CFT deformed by $\phi^{2n}$ interaction, 
which will be called the generalized Wilson-Fisher CFTs \cite{Gliozzi:2016ysv,Gliozzi:2017hni,Gliozzi:2017gzh}.

The multiplet recombination approach initiated in \cite{Rychkov:2015naa} 
has been extended in various aspects 
and achieved considerable success in different CFTs \cite{Basu:2015gpa,Yamaguchi:2016pbj,Hasegawa:2016piv,Ghosh:2015opa,Raju:2015fza,Gliozzi:2016ysv,Roumpedakis:2016qcg,Soderberg:2017oaa,Behan:2017emf,Gliozzi:2017hni,Gliozzi:2017gzh,Nishioka:2022odm,Nishioka:2022qmj,Antunes:2022vtb}. 
See also \cite{Safari:2017irw, Nii:2016lpa,Hasegawa:2018yqg,Skvortsov:2015pea,Giombi:2016hkj,Giombi:2017rhm,Giombi:2016zwa,Codello:2017qek,Codello:2018nbe,Antipin:2019vdg,Vacca:2019rsh,Giombi:2020rmc,Giombi:2020xah, Dey:2020jlc,Giombi:2021cnr,Safari:2021ocb,Zhou:2022pah,Bissi:2022bgu,Herzog:2022jlx,Giombi:2022vnz,SoderbergRousu:2023pbe} for the closely related studies with more emphasis on the equations of motion from 
concrete Lagrangians, which is sometimes called the Dyson-Schwinger or Schwinger-Dyson equations.\footnote{The division may not be completely definite, as some results can be derived from both perspectives. }
At the moment, only leading order corrections have been derived from this approach in the literature. 
It was suggested in \cite{Rychkov:2015naa} that it may be useful to consider four-point functions. 
This is because the second consistency requirement mentioned earlier leads to nontrivial crossing constraints. 
In fact, the crossing constraints with some spectral assumptions are expected to be strong enough to determine the deformed data by themselves,  
based on the nonperturbative numerical results \cite{ElShowk:2012ht, El-Showk:2014dwa, Kos:2014bka, Kos:2016ysd,Poland:2018epd} and the perturbative analytic results \cite{Alday:2017zzv,Henriksson:2018myn}. 

In this work, we will derive new results for the leading terms of the broken higher-spin currents in the higher-derivative generalization of multicritical theories. 
The more standard theories, such as $k=1$ or $n=2,3$, are covered as special cases of our general formulas. 
As a necessary step of merging the two consistency-requirement approaches for the generalized Wilson-Fisher CFTs, 
we use the Lorentzian inversion formula \cite{Caron-Huot:2017vep} to verify that our general results from the multiplet recombination method are compatible with 
the crossing constraints.\footnote{In principle, some operators could have vanishing OPE coefficients in the Gaussian limit, 
so they only exist in the deformed CFT, such as the evanescent operators at noninteger $d$ \cite{Collins:1984}.  
Constraints on subleading terms can be derived from the absence of new low-lying states in the crossing solutions  
related to the mixed OPEs in the multiplet recombination 
\cite{subleading-null}. 
The decoupling requirements can be viewed as the null state conditions 
from the null bootstrap perspective \cite{Li:2022prn,Li:2023nip}. 
Furthermore, there may exist a correspondence between null states in the interacting theory and in the free limit, such as the free and interacting equations of motion, 
which can lead to more nontrivial constraints. 
}

In Sec. \ref{sec:Partially conserved current}, we give an introduction to higher-spin symmetries and partially conserved currents.
In Sec. \ref{sec:Embedding formalism}, we briefly review the embedding formalism, 
especially the case of scalar-scalar-(spin $\ell$) three-point function. 
In Sec. \ref{sec:Anomalous dimensions of broken partially conserved currents}, 
we derive the anomalous dimensions of broken higher-spin currents using the conformal multiplet recombination, then we use the Lorentzian inversion formula to verify the consistency with crossing symmetry. 
In Sec. \ref{sec:O(N) models}, the results are extended to the O($N$) models.
In Appendix \ref{sec:b}, we provide the leading order expressions of some OPE coefficients involving $\phi^{2n-2}$, $\phi^{2n}$, and the O($N$) generalization of the former one.
In Appendix \ref{sec:Ratio of 3 point function coefficients}, we present the calculation of the ratios of three-point function coefficients.
The expressions of the light cone expansion of conformal blocks are given in Appendix \ref{Lightcone expansion of conformal blocks}.
We provide some details of the inversion procedure at subleading twist, and sub-subleading twist in Appendix \ref{Subleading twist and sub-subleading twist}.

\section{Partially conserved currents and symmetries}
\label{sec:Partially conserved current}
According to Noether's theorem, 
a Noether current satisfying $\pa^\m J_\m=0$ is associated with 
a linearly realized global symmetry of the Lagrangian. 
For the standard free scalar CFT with $k=1$, the symmetric traceless bilinear primary operators of 
the schematic form $\mathcal {J}=\phi\,\pa^\ell\phi$ are higher-spin conserved currents 
with twist $\t\equiv\D-\ell=d-2$:
\begin{align}
\pa^{\m_1}\mathcal {J}_{\m_1\m_2\dots \m_\ell}=0\,.
\label{conservation-law}
\end{align}
Noether currents can be obtained by contracting the higher-spin conserved currents with the conformal Killing tensors
\begin{align}
J_{\m}^{(\ell)}=\mathcal{J}_{\m\m_1\dots \m_{\ell-1}}\zeta^{\m_1\dots \m_{\ell-1}}\,,
\end{align} 
where $\zeta^{\m_1\dots \m_{\ell-1}}$ is symmetric and traceless.\footnote{The conformal Killing equation is
\begin{align}
\pa^{(\m_1}\zeta^{\m_2\dots \m_\ell)_T}=0\,,
\end{align}
where $T$ indicates the traceless part, 
so this is a conformal version of the usual Killing equation. }
One can show that the Lagrangian is invariant up to a total derivative 
under the symmetry transformation
\begin{align}
\d_\zeta\phi=\zeta^{\m_1\dots \m_{\ell-1}}\pa_{\m_1}\dots\pa_{\m_{\ell-1}}\phi+\dots\,,
\end{align}
where the last ellipsis indicates other terms with derivatives acting on $\zeta$. 
For instance, the spin-2 conserved current, i.e., the stress tensor, is associated with the global conformal transformation $\d \phi=\zeta^\m\pa_\m\phi+\frac{\D_\phi}{d}(\pa_\m \zeta^\m) \phi$.

For $k>1$, we can consider higher-derivative primary bilinear operators of the schematic form $\mathcal {J}_\ell^{(m)}=\phi\,\pa^\ell\Box^m\phi$ with twist $\t\equiv \D-\ell=d-2k+2m$, 
where $m=0,1,\ldots,k-1$ is related to the number of contracted derivative-indices.
In principle, the explicit expressions of $\mathcal {J}^{(m)}_{\ell}$ are determined by the conditions that they are primary and symmetric traceless.\footnote{ An operator $\mathcal{O}$ is primary if it satisfies $[K_{\m},\mathcal{O}(0)]=0$, where $K_{\m}$ are the generators of special conformal transformations.
In some peculiar cases of the free $\Box^k$ CFT, there exist operators that are neither primary nor descendant \cite{Brust:2016gjy}, which occurs only when $d=3,5,\ldots,2k-1$ or $d=2k+2,2k+4,\ldots,4k-2$.
In this work, we consider the upper critical dimension at $d_{\text{u}}=2nk/(n-1)>2k$.
To be in a peculiar case, $d_{\text{u}}$ should be an even integer.
Since $n$ and $n-1$ are coprime, $d_{\text{u}}$ is even only if $n=2$, or $k$ is a multiple of $n-1$.
The case $n=2$ implies $d_{\text{u}}=4k$, and this is outside the range of the peculiar cases.
The other possibility is ruled out by our assumption that $k$ and $n-1$ have no common divisor. 
Therefore, the peculiar cases do not appear in our discussion. } 
We believe that $\mathcal{J}^{(m)}_{\ell}$ are nondegenerate, but we do not have a general proof.\footnote{ Following \cite{Penedones:2010ue,Fitzpatrick:2011dm,Bekaert:2015tva}, 
we solve for the explicit expressions of $\mathcal {J}^{(m)}_{\ell}$ using the primary and the symmetric traceless conditions.
For generic $\Delta_\phi$ and $\ell$, we can determine $\mathcal {J}^{(m)}_{\ell}$ at specific $m$.
The solution at each $m$ is unique up to normalization, so there seems no degeneracy in $\mathcal{J}^{(m)}_{\ell}$.
We have checked this for high values of $m$, and we believe that $\mathcal {J}^{(m)}_{\ell}$ are not degenerate.
In Appendix D of \cite{Bekaert:2015tva}, the conjectured expression for the general solution of  $\mathcal{J}^{(m)}_{\ell}$ is unique up to normalization. }

Intuitively, the $m>0$ trajectories do not vanish automatically 
because the equation of motion $\Box^k \phi=0$ is of higher derivatives.\footnote{ Together with $[K_{\m},\mathcal {J}^{(m)}_{\ell}(0)]=0$ and the symmetric traceless condition, the equation of motion $\Box^k \phi=0$ implies that $\mathcal {J}^{(m)}_{\ell}$ vanishes for $m>k-1$. }
The highest trajectory with $m=k-1$ has twist $\t=d-2$ 
and corresponds to the usual conserved higher-spin currents satisfying \eqref{conservation-law}. 
For $m<k-1$, they are the partially conserved higher-spin currents satisfying
\begin{align}\label{partial-conservation}
\partial^{\mu_{1}}\ldots\partial^{\mu_{c}}
\mathcal{J}^{(m)}_{\mu_{1}\ldots\mu_{\ell}}=0\,,
\quad \ell\geqslant c\equiv2(k-m)-1\,,
\end{align}
which do not vanish if the number of contracted derivatives is less than $c$.
They are nonunitary as their twists violate the unitarity bound, i.e., $\t<d-2$. 
In addition, the operators with spin lower than $c$ do not satisfy the partial conservation laws. 
A concrete example with $k=2$ is the triply conserved current
\begin{align}
\mathcal{J}_{\mu\nu\rho\sigma}^{(m=0)}=
\[\frac{(d-4)(d-2)}{3d(d+2)}\phi\partial_{\mu}\partial_{\nu}\partial_{\rho}\partial_{\sigma}\phi-\frac{4(d-2)}{3d}\partial_{\mu}\phi\partial_{\nu}\partial_{\rho}\partial_{\sigma}\phi+\partial_{\mu}\partial_{\nu}\phi\partial_{\rho}\partial_{\sigma}\phi\]_\text{ST}\,,
\end{align}
where ST indicates symmetric and traceless projection. 
This spin-4 current vanishes when contracted with $3$ derivatives on the equation of motion $\Box^{2}\phi=0$.

One can also construct Noether currents by contracting the partially conserved currents with 
higher-order generalization of the conformal Killing tensors\footnote{The generalized conformal Killing equation is   
\begin{align}
\pa^{(\m_1}\dots\pa^{\m_c}\zeta^{\m_{c+1}\dots\m_\ell)_T}=0\,.
\end{align}}
\begin{align}
J_{\m}^{(\ell,c)}=\;&\sum_{i=0}^{c-1}(-1)^i\pa^{\m_1}\dots\pa^{\m_{i}}\mathcal{J}^{(m)}_{\m\m_1\m_2\dots \m_{\ell-1}}
\pa^{\m_{i+1}}\dots\pa^{\m_{c-1}}\zeta^{\m_{c}\dots \m_{\ell-1}}\,.
%\nn=\;&\mathcal{J}^{(m)}_{\m\m_1\m_2\dots \m_{\ell-1}}
%\pa^{\m_1}\dots\pa^{\m_{c-1}}\zeta^{\m_{c}\dots \m_{\ell-1}}-\pa^{\m_1}\mathcal{J}^{(m)}_{\m\m_1\m_2\dots \m_{\ell-1}}\pa^{\m_{2}}
%\dots\pa^{\m_{c-1}}\zeta^{\m_{c}\ldots \m_{\ell-1}}
%\nn
%&+\pa^{\m_1}\pa^{\m_{2}}\mathcal{J}^{(m)}_{\m\m_1\m_2\dots \m_{\ell-1}}\pa^{\m_{3}}
%\dots\pa^{\m_{c-1}}\zeta^{\m_{c}\dots \m_{\ell-1}}
%+\dots
%\nn
%&+(-1)^{c-1}\pa^{\m_1}\dots \pa^{\m_{c-1}}\mathcal{J}^{(m)}_{\m\m_1\m_2\dots \m_{\ell-1}}\zeta^{\m_{c}\dots \m_{\ell-1}}\,.
\end{align}
The corresponding symmetry transformation reads
\begin{align}
\d_\zeta\phi=\zeta^{\m_1\dots \m_{\ell-c}}\pa_{\m_1}\dots\pa_{\m_{\ell-c}}\Box^{\frac{c-1}{2}}\phi+\dots\,,
\end{align}
where the last ellipsis indicates other terms with different derivative contractions and they are determined by the invariance of the Lagrangian up to a total derivative.

The global symmetries should form a closed algebra. 
The commutator of two transformations should be associated with 
a linear combination of certain Killing tensors
\begin{align}
\d_{\zeta_1}\d_{\zeta_2}-\d_{\zeta_2}\d_{\zeta_1}=\d_{[\zeta_1,\zeta_2]}+\text{(trivial)}\,,
\end{align}
then the global symmetry leads to a Lie algebra structure. 
This is the nontrivial symmetries of the equation of motion $\Box^k\phi=0$, 
which generalizes the $k=1$ higher-spin algebra $hs_1$ to the higher-order counterpart $hs_k$ \cite{Eastwood-Leistner,Eastwood:2002su,Gover:2009,Michel,Bekaert:2013zya,Joung:2015jza}.

At (generalized) Wilson-Fisher fixed points, most of the (partial-)conservation laws are broken 
and the corresponding currents acquire anomalous dimensions.\footnote{In the AdS/CFT correspondence, 
the bulk fields become massive due to a Higgs-like mechanism \cite{Bianchi:2005ze}. } 
We would like to compute their leading corrections using CFT methods. 

\section{Embedding formalism}
\label{sec:Embedding formalism}
In this section, we will give a brief introduction to the embedding formalism 
in which the conformal transformations are linearly realized
and thus the consequences of conformal symmetry can be readily deduced. 
To avoid the complication of explicit tensor structures,
we will apply the index-free notation to the correlators involving spinning operators 
by contracting them with auxiliary polarization vectors. 

For $d>2$, the conformal symmetry is finite-dimensional. 
The symmetry generators are associated with the Poincar\'e transformations, dilatation, and special conformal transformations. 
The first two kinds of symmetries mean that a physical field is labeled by spin\footnote{We will focus on bosonic fields of the symmetric traceless type and will not consider anti-symmetric or mixed-symmetry fields.  } 
and scaling dimension. 
However, the implications of special conformal transformations are less transparent, 
especially for spinning fields, 
due to the fact that they are not realized linearly.\footnote{One may view a special conformal transformation as the ``conjugate'' of a translation, 
which can be produced by a composition of inversion, translation, and inversion. 
Although the translation symmetries are not realized linearly, 
their consequences are easy to understand. 
}

The conformal group of a Euclidean conformal field theory is SO($d+1,1$),  
so it is natural to consider the Minkowski space $\mathbb{R}^{d+1,1}$ as first proposed by Dirac \cite{Dirac:1936fq}. 
The physical space $\mathbb{R}^{d}$ of $d$ dimensions should be a subspace of 
this $(d+2)$-dimensional embedding space.\footnote{The embedding formalism has also been used in the study of higher-spin fields in de Sitter and anti de Sitter space \cite{Fronsdal:1978vb,Biswas:2002nk}, 
which is also called the ambient space formalism (see \cite{Bekaert:2004qos} for example).}
The use of the embedding space in the study of conformal field theory has a long history \cite{Mack:1969rr,Boulware:1970ty,Ferrara:1973yt,Ferrara:1973eg}. 
At the price of introducing two more dimensions and potentially redundant degrees of freedom, 
the conformal group is now realized as the Lorentz group of linear isometries and thus 
their implications become more manifest.

\begin{figure}[h]
	\centering
	\includegraphics[width=.5\textwidth,origin=c,angle=0]{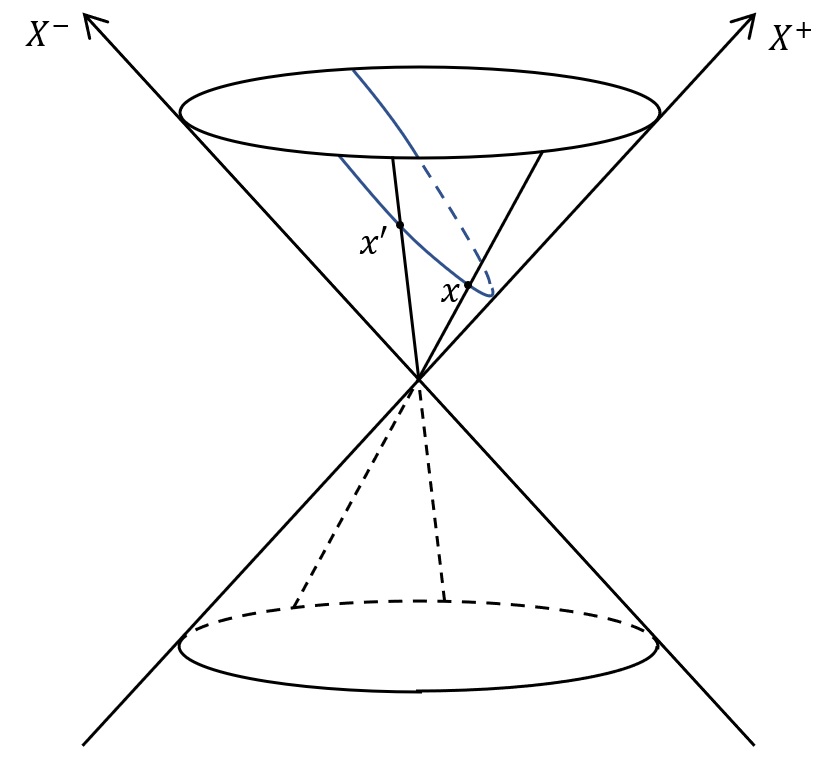}
	\caption{In the light cone of the embedding space, the physical space is associated with the Poincar\'e section. 
	An SO($d+1,1$) transformation maps the light ray associated with the physical point $x$ to that of another physical point $x'$. \label{embedding} }
\end{figure}

We will follow the discussions in \cite{Costa:2011mg,Rychkov:2016iqz} and refer to them for more details. 
We use uppercase letters to indicate embedding space objects and lowercase letters for those of physical space. 
To reduce two dimensions, 
we can consider the light cone of the embedding space and identify the physical space as the Poincar\'e section. 
The light cone coordinates of $\mathbb{R}^{d+1,1}$ are
\begin{align}
	X^{A}=(X^{+},X^{-},X^{a})\,,
\end{align}
and the metric $\eta_{AB}$ is given by
\begin{align}
	\mathrm{d}S^{2}=\eta_{AB}X^AX^B=-\mathrm{d}X^{+}\mathrm{d}X^{-}+\sum_{a=1}^{d}(\mathrm{d}X^a)^2\,. 
\end{align}
The Poincar\'e section is given by $X^+=1$ whose coordinates are
\begin{align}
	X_{\text{Poincar\'e}}^{A}=(1,x^{2},x^{a})\,.
\end{align}
As shown in Fig. \ref{embedding}, a point $x$ in the Poincar\'e section lies in a null ray $X^A=\l(1,x^{2},x^{a})$ in the embedding space, and $g\in \text{SO}(d+1,1)$ transforms this light ray to the light ray that meets the Poincar\'e section at $x'$. 
In this way, we can define the action of $g$ on a physical point $x$ by the Lorentz transformations of the light rays. 
This is indeed a conformal transformation as it induces a local rescaling of the physical metric, 
which is associated with the change in $\l$. 

The next step is to extend the transformations of fields to the embedding space counterparts. 
We first need to uplift the physical fields to the light cone. 
This provides a more economical way to derive the constraints of conformal symmetry on correlation functions. 
There is a correspondence between a symmetric traceless primary field $f_{a_{1},...,a_{\ell}}(x)$ with spin-$\ell$ 
and scaling dimension $\D$ 
and a symmetric, traceless, homogeneous, and transverse SO($d+1,1$) tensor $F_{A_{1},\ldots,A_{\ell}}(X)$ 
defined on the light cone. 
An embedding tensor of homogeneity $-\D$ satisfies 
\begin{align}
F_{A_{1},\ldots,A_{\ell}}(\l X)=\l^{-\D}F_{A_{1},\ldots,A_{\ell}}(X)\,,
\end{align} for $\l>0$. 
The transversality condition reads 
\begin{align}
X^{A_{i}}F_{A_{1},\ldots,A_{i},\ldots,A_{\ell}}=0\,.
\end{align} 
The homogeneity condition extends the definition of fields in the Poincar\'e section to the complete light cone. 
The transversality constraint eliminates one redundant component for each index. 
We can recover the physical tensor by projecting $F$ onto the physical space 
\begin{align}\label{Proj}
	f_{a_{1},\ldots,a_{\ell}}(x)=\frac{\pa X^{A_{1}}_{\text{Poincar\'e}}}{\pa x^{a_{1}}}\ldots\frac{\pa X^{A_{\ell}}_{\text{Poincar\'e}}}{\pa x^{a_{\ell}}}F_{A_{1},\ldots,A_{\ell}}(X_{\text{Poincar\'e}})\,.
\end{align}
This projection gives us a symmetric traceless field $f$. The symmetric property follows immediately from the fact that $F$ is symmetric. The traceless property comes from the tracelessness and transversality of $F$. 
Some redundant degrees of freedom are removed by the fact that terms proportional to $X^{A_i}$ are orthogonal to the projection vector 
due to $X^2=0$. 
Finally, the projection \eqref{Proj} of $F$ under SO($d+1,1$) transformations 
produce the correct transformations of a symmetric traceless primary field of spin-$\ell$ \cite{Ferrara:1973eg,Weinberg:2010fx}. 

In the index-free notation, 
the index structure of a symmetric traceless tensor $f_{a_{1},...,a_{\ell}}(x)$ is encoded in a polynomial of polarization vector $z$
\begin{align}
	f(x,z)=f_{a_{1},\ldots,a_{\ell}}(x)\,z^{a_{1}}\ldots z^{a_{\ell}}\,,
\end{align}
where $z\in\mathbb{C}^{d}$ and we can set $z^{2}=0$ because $f_{a_{1},\ldots,a_{\ell}}$ is symmetric traceless.\footnote{The symmetric traceless tensor $f_{a_{1},\ldots,a_{\ell}}$ can be recovered from $f(x,z)|_{z^2=0}$.}
In the embedding space, $F_{A_{1},\ldots,A_{\ell}}(X)$ can be encoded by a polynomial
\begin{align}
	F(X,Z)=F_{A_{1},\ldots,A_{\ell}}(X)\,Z^{A_{1}}\ldots Z^{A_{\ell}}\,,
\end{align}
where $Z\in\mathbb{C}^{d+2}$. We can set $Z^{2}=0$ and $Z\cdot X=0$ because $F$ is symmetric, traceless, and transverse.  

In the present work, it suffices to consider the scalar-scalar-(spin-$\ell$) three-point function 
\begin{align}
\langle F_{1}(X_{1})F_{2}(X_{2})F_3(X_{3},Z_{3})\rangle\,.
\end{align}  
Since $Z_3^2=0$ and $Z_3\cdot X_3=0$, we know that $Z_3$ only appears in $Z_3\cdot X_1$ and $Z_3\cdot X_2$. 
It turns out that the transversality condition leads to the following building block $(Z_{3}\cdot X_{1})(X_{2}\cdot X_{3})-(Z_{3}\cdot X_{2})(X_{1}\cdot X_{3})$.
Together with SO($d+1,1$) invariance, homogeneity, and spin-$\ell$ conditions, the three-point function is fixed up to a constant
\begin{align}
	\Scale[0.95]{ 
	\langle F_{1}(X_{1})F_{2}(X_{2})F_3(X_{3},Z_{3})\rangle= }
	\Scale[1.1]{ 
	\;\frac{\text{constant}\big((Z_{3}\cdot X_{1})(X_{2}\cdot X_{3})-(Z_{3}\cdot X_{2})(X_{1}\cdot X_{3})\big)^{\ell}}{(X_{1}\cdot X_{2})^{\frac{\D_{1}+\D_{2}-\D_{3}+\ell}{2}}(X_{1}\cdot X_{3})^{\frac{\D_{1}+\D_{3}-\D_{2}+\ell}{2}}(X_{2}\cdot X_{3})^{\frac{\D_{2}+\D_{3}-\D_{1}+\ell}{2}}} }\,.
\end{align}
As the embedding and physical coordinates are related by
\begin{equation}
	X_{i}\cdot X_{j}=-\frac{1}{2}(x_{i}-x_{j})^{2},\qquad Z\cdot X_{i}=z\cdot(x_{i}-x_{3})\,,
\end{equation}
the three-point function in the physical space reads
\begin{align}
	\< \mathcal O_{1}(x_{1})\mathcal O_{2}(x_{2})\mathcal O_{3}(x_{3},z)\>=\frac{\text{constant}\left(z\cdot\left(\frac{x_{13}}{x_{13}^{2}}-\frac{x_{23}}{x_{23}^{2}}\right)\right)^{\ell}}{x_{12}^{\D_{1}+\D_{2}-\D_{3}+\ell}x_{13}^{\D_{1}+\D_{3}-\D_{2}-\ell}x_{23}^{\D_{2}+\D_{3}-\D_{1}-\ell}}\,,
	\label{3-point-spinning}
\end{align}
where $x_{ij}\equiv|x_{i}-x_{j}|$. 
Below, we will consider the action of Laplacians with respect to $x_1$, $x_2$ on this three-point function. 
Despite the simplicity of the index-free notation, 
the action of higher-order Laplacians on the three-point function can lead to complicated expressions. 
Inspired by \cite{Gliozzi:2017gzh}, we simplify the functional form of the three-point function 
by focusing on the leading term in the limit $x_{3}\rightarrow\infty$:
\begin{align}
	\< \mathcal O_{1}(x_{1})\mathcal O_{2}(x_{2})\mathcal O_3(x_{3},z)\>=\frac{\text{constant}\left(z\cdot x_{12}\right)^{\ell}}{x_{12}^{\Delta_{1}+\Delta_{2}-\Delta_{3}+\ell}x_{3}^{2\Delta_{3}}}+O\left(x_3^{-(2\D_\ell+1)}\right)\,.
	\label{3-point-spinning-simplified}
\end{align}
Then it is straightforward to derive the general results of higher-order Laplacians acting on \eqref{3-point-spinning-simplified}.
In the radial quantization, the $x_3\rightarrow \infty$ limit means the out state is given by $\langle \mathcal O_3|=\lim_{x\rightarrow \infty}x_3^{2\D_{\mathcal O}}\langle 0|\mathcal O_3(x_3)$, 
so we extract the $\mathcal O_3(x_1)$ contribution in the OPE $\mathcal O_1(x_1)\times\mathcal O_2(x_2)$ to the in state.

\section{Anomalous dimensions of partially conserved currents}
\label{sec:Anomalous dimensions of broken partially conserved currents}
In this section, we study the anomalous dimensions of partially conserved currents $\mathcal{J}^{(m)}_{\ell}$ in the $\phi^{2n}$ deformation of the free $\Box^k$ CFT. We first use the multiplet recombination method to deduce the general results, 
then we examine the results using the analytic conformal bootstrap.
As mentioned in Sec. \ref{sec:Introduction}, we will assume that $k$ and $n-1$ have no common divisor, so we do not need to consider derivative interactions.

The Lagrangian formulation of the $\phi^{2n}$ theory in $d=d_\text{u}-\e$ reads
\begin{align}
S\propto\int\mathrm{d}^{d}x\;\left(\phi\,\Box^{k}\phi+g\m^{(n-1)\e}\phi^{2n}\right)\,,
\end{align}
where the upper critical dimension is $d_\text{u}=2nk/(n-1)$. 
We are interested in the CFT describing the IR fixed point of the RG flow 
triggered by $\phi^{2n}$. 
The IR CFT itself can be thought of as a consistent deformation of the free CFT, 
parametrized by the dimension $d$. 

\subsection{Multiplet recombination}
\label{sec:Multiplet recombination}
In a smooth deformation, the free CFT states should extend to the deformed CFT. 
On the other hand, the scaling dimensions and three-point function coefficients of the deformed operators 
can change as smooth functions of $\e$,  
which should reduce to the free values in the limit $\e\rightarrow 0$. 
For notational simplicity, we use $\mathcal O_\text{f}$ to denote the free CFT operator 
and  $\mathcal O$ the corresponding deformed CFT operator, 
without adding the subscript WF. 
The deformed operators can be interpreted as the renormalized operators in the Lagrangian formulation.

As discussed in the introduction, the lowest scalar primary $\phi_\text{f}$ in the free CFT satisfies the equation of motion
\begin{align}
\Box^k\phi_\text{f}=0\,.
\end{align}
From the CFT perspective, the null state condition on the descendant $\Box^k\phi_\text{f}$ 
is a consequence of conformal symmetry and $\D_{\phi_\text{f}}=d_\text{u}/2-k$, 
without referring to a Lagrangian description. 
In the interacting CFT, it is expected that the corresponding operator $\phi$ acquires an anomalous dimension 
\begin{align}
\g_\phi=\D_\phi-d/2+k\,,
\end{align} 
so the descendant $\Box^k\phi\neq0$ is a physical state at $d=d_\text{u}-\e$. 
All the physical operators in the free CFT should have interacting counterparts.\footnote{However, some operators in the deformed CFT can decouple in the Gaussian limit. }
Furthermore, we assume that $\Box^k\phi$ corresponds to a physical operator in the Gaussian limit,  
but we need to change the normalization of $\Box^k\phi$
to obtain a finite-norm state. 
By $\phi^{2n}$ deformation, we mean that the Gaussian limit of $\Box^k\phi$ corresponds to $\phi_\text{f}^{2n-1}$
\begin{align}\label{rec}
\lim_{\e\rightarrow 0} \a^{-1}\Box^k\phi=\phi_\text{f}^{2n-1}\,,
\end{align}
where $\a=\a(\e)$ is a function of $\e$ with $\lim_{\e\rightarrow0}\a=0$, i.e., 
we introduce a singular change in the normalization of $\Box^k\phi$ to obtain a finite-norm state. 
In other words, we identify a descendant of $\phi$ with the deformed operator of $\phi_\text{f}^{2n-1}$
\begin{align}
\Box^k\phi= \a\,\phi^{2n-1}\,.
\end{align}
One can check that the scaling dimensions match the Gaussian limit, i.e.,  
$\lim_{\e\rightarrow 0}\D_{\phi}+2k=\D_{\phi_\text{f}}+2k=(2n-1)\D_{\phi_\text{f}}$. 
Although $\phi_\text{f}^{2n-1}$ is a primary in the free theory, its deformed version is a descendant of $\phi$. 
In this sense, the free multiplets $\{\phi\}_\text{free}$ and $\{\phi^{2n-1}\}_\text{free}$ recombine 
to form the Wilson-Fisher multiplet $\{\phi\}_\text{WF}$. 

To determine the leading behavior of $\a(\e)$, 
let us examine the two-point function of $\Box \phi$. 
As an illustrative example, we consider the standard case of $k=1,n=2$
\begin{align}
	\langle \Box \phi(x_1)\Box\phi(x_2)\rangle
	&=\Box_{x_1}\Box_{x_2}\langle  \phi(x_1)\phi(x_2)\rangle
	\nn
	&=16\D_{\phi}(\D_{\phi}+1)(\D_{\phi}-d/2+1)(\D_{\phi}-d/2+2)\frac{\l_{\phi\phi\mathds{1}}}{x_{12}^{2\D_\phi+4}}
	\nn
	&=32\g_\phi (1+\dots)\frac{\l_{\phi\phi\mathds{1}}}{x_{12}^{2\D_\phi+4}}
\,,
\end{align}
where $\lim_{\e\rightarrow 0}\D_\phi=\D_{\phi_\text{f}}=1$ and the ellipses indicate subleading terms in $\e$. 
The two-point function coefficient $\l_{\phi\phi\mathds{1}}$ is finite in the limit $\e\rightarrow 0$.  
We have used the identity 
\begin{align}\label{Box-2p}
\Box_{x_1}\left (\frac{1}{x_{12}^{2\D}}\right)=4\D(\D-d/2+1)\frac{1}{x_{12}^{2(\D+1)}}\,.
\end{align}
Therefore, the finite-norm condition of $\a^{-1}\Box\phi$ implies
\begin{align}\label{regular-limit}
\a \propto (\g_\phi)^{1/2}\left(1+\dots\right)\,,
\end{align}
where the proportionality factor should be finite as $\e\rightarrow 0$. 
In the Gaussian limit, the Wilson-Fisher correlator reduces to the Gaussian correlator 
\begin{align}
\lim_{\e\rightarrow 0}\a^{-2}\langle \Box \phi(x_1)\Box\phi(x_2)\rangle=\langle \phi^3_\text{f}(x_1)\phi^3_\text{f}(x_2)\rangle=\frac{\l_{\phi_\text{f}^3\phi_\text{f}^3\mathds{1}}}{x_{12}^{2\D_{\phi_\text{f}}}}\,,
\end{align} 
so we have
\begin{align}\label{2box-2p-match}
\lim_{\e\rightarrow0} 32\,\a^{-2} \g_\phi =\frac{\l_{\phi_\text{f}^3\phi_\text{f}^3\mathds{1}}}{\l_{\phi_\text{f}\phi_\text{f}\mathds{1}}}\,. 
\end{align}
where $\lim_{\e\rightarrow 0}\l_{\phi\phi \mathds{1}}=\l_{\phi_\text{f}\phi_\text{f}\mathds{1}}$ is used.
The matching condition in the Gaussian limit only determines the leading behavior of $\a$ in the $\e$ expansion, 
not the full functional form of $\a(\e)$, 
as the choice of normalization is not fixed at subleading orders in $\e$. 
From this simple example \eqref{2box-2p-match}, 
we can see the general structure of matching conditions: 
\begin{itemize}
\item
The left-hand side involves the Gaussian limit of a combination of $\a$ and scaling dimensions. 
The action of $\Box^k$ leads to Pochhammer symbols.  
Furthermore, a factor needs more care if its Gaussian limit vanishes. 
\item
The right-hand side is given by a ratio of Gaussian OPE coefficients. 
If the OPE coefficient in the denominator vanishes in the Gaussian limit, 
then we should move it back to the left-hand side. 
\end{itemize}

In a systematic analysis of the leading-order matching conditions, 
it is sufficient to consider only two- and three-points functions, as they encode all the local CFT data. 
The WF correlators should reduce to the free ones in the Gaussian limit $\epsilon\rightarrow0$:
\begin{align}
	\lim_{\e\rightarrow 0}\<\mathcal{O}_{1}\mathcal{O}_{2}\ldots\>
	=\<\mathcal{O}_{1,\text{f}}\,\mathcal{O}_{2,\text{f}}\ldots\>\,.
\end{align}
For the free theory operators, 
we use the normalization that the three-point function coefficients $\l_{\mathcal{O}_{1,\text{f}}\mathcal{O}_{2,\text{f}}\mathcal{O}_{3,\text{f}}}$ are given by Wick contractions, 
so the two-point function coefficients of composite operators will be different from one. 
For composite operators associated with multiple $\phi$ and derivatives, 
we assume that the operators under consideration are nondegenerate, 
otherwise we need to first solve the mixing problem to derive more useful constraints. 
For a deformed operator of the schematic form $\mathcal O=\pa^{i_1}\phi_\text{f}^{i_2}$ in the Gaussian limit, 
the anomalous dimension is defined as
\begin{align}\label{gamma-definition}
\g_{\mathcal O}=\D_{\mathcal O}-i_1-i_2(d/2-k)\,.
\end{align}
Since $\D_{\phi^{2n-1}}=\D_{\Box^k\phi}=\D_\phi+2k$, there is a relation between the anomalous dimensions of $\phi,\phi^{2n-1}$
\begin{align}\label{gamma-descendant}
\g_{\phi^{2n-1}}=(n-1)\e+\g_\phi\,.
\end{align}
For general $k$, the matching condition of the two-point function of the descendant $\Box^k\phi$ is 
\begin{align}\label{BoxBox2P}
	\lim_{\e\rightarrow0}\alpha^{-2}\<\Box^{k}\phi(x_{1})\,\Box^{k}\phi(x_{2})\>&
	=\<\phi_\text{f}^{2n-1}(x_{1})\,\phi_\text{f}^{2n-1}(x_{2})\>\,,
\end{align}
Using \eqref{Box-2p}, one can verify that the functional dependence on $x_i$ matches. 
Then the matching of coefficients gives
\begin{align}
	\lim_{\e\rightarrow 0}\,\a^{-2}\,4^{2k}(\D_{\phi})_{2k}\,(\D_{\phi}-d/2+1)_{2k}
	=\frac{\l_{\phi^{2n-1}_\text{f}\phi^{2n-1}_\text{f}\mathds{1}}}{\l_{\phi_\text{f}\phi_\text{f}\mathds{1}}}\,.
\end{align}
Since $\g_\phi$ vanishes in the limit $\e\rightarrow 0$, we can omit some subleading terms. 
Substituting with the free theory data, we obtain
\begin{align}\label{match-1}
4^{2k}\(\frac{k}{n-1}\)_{2k}(1-k)_{k-1}k!\,\lim_{\e\rightarrow 0}\left(\a^{-2}\g_\phi\right)=(2n-1)!\,,
\end{align}
where $(x)_{y}={\G(x+y)}/{\G(x)}$ is the Pochhammer symbol and $\Gamma(x)$ is the Gamma function. 
Note that the factor $\g_\phi=\D_\phi-d/2+k$ comes from $(\D_{\phi}-d/2+1)_{2k}$. 

Then we consider the three-point functions with finite $\l_{ijk}$ in the Gaussian limit on both sides\footnote{ There is only one way to construct an operator of the form $\phi^{p}$, so we do not need to worry about degeneracies. }
\begin{align}\label{Box3P}
	\lim_{\e\rightarrow 0}\a^{-1}\<\Box^{k}\phi(x_{1})\phi^{p}(x_{2})\phi^{p+1}(x_{3})\>=
	\<\phi_\text{f}^{2n-1}(x_{1})\phi_\text{f}^{p}(x_{2})\phi_\text{f}^{p+1}(x_{3})\>\,.
\end{align}
For $p\neq2n-1,2n-2$, the left-hand side can be derived from the action of Laplacians on the correlator of primary operators. 
Using \eqref{3-point-spinning-simplified} and \eqref{Box-2p}, we obtain
\begin{align}\label{match-2}
	2^{2k-1}{(k-1)!}\left(1-\frac{n k}{n-1}\right)_{k}
	\lim_{\e\rightarrow 0}\left(\a^{-1}\left(\g_{1}^{\phantom{()}\!\!}+\g_{p}^{\phantom{()}\!\!}-\g_{p+1}^{\phantom{()}\!\!}\right)\right)
	=\frac{\l_{2n-1,p,p+1}}
	{\l_{1,p,p+1}}\,,
\end{align}
where $\g_{p}^{\phantom{()}\!\!}$ is the anomalous dimension of $\phi^{p}$, and $\l_{p_1,p_2,p_3}$ is the free three-point function coefficient of $\phi^{p_1}_{\text{f}}$, $\phi^{p_2}_{\text{f}}$, and $\phi^{p_3}_{\text{f}}$:
\begin{align}
	\g_{p}^{\phantom{()}\!\!}\equiv\g_{\phi^{p}}\,,\qquad \l_{p_1,p_2,p_3}\equiv\l_{\phi^{p_1}_{\text{f}}\phi^{p_2}_{\text{f}}\phi^{p_3}_{\text{f}}}\,.
\end{align}
Here the parameter $\D$ in \eqref{Box-2p} is given by $(\Delta_{\phi}+i+\Delta_{\phi^{p}}-\Delta_{\phi^{p+1}})/2$.  
For $p=2n-2,2n-1$, the correlator involves the descendant $\phi^{2n-1}=\a^{-1}\Box^{k}\phi$,  
so we consider the matching conditions involving one more $\phi$
\begin{align}
	\label{boxk-1-1-2n}
\lim_{\e\rightarrow 0}\a^{-1}\<\Box^{k}\phi(x_{1})\,\phi(x_{2})\,\phi^{2n'}(x_{3})\>=
	\<\phi_\text{f}^{2n-1}(x_{1})\,\phi_\text{f}(x_{2})\,\phi_\text{f}^{2n'}(x_{3})\>\,,
\end{align}
and
\begin{align}\label{boxk-boxk-1-1-2n}
\lim_{\e\rightarrow 0}\a^{-2}\<\Box^{k}\phi(x_{1})\,\Box^{k}\phi(x_{2})\,\phi^{2n'}(x_{3})\>=
	\<\phi_\text{f}^{2n-1}(x_{1})\,\phi_\text{f}^{2n-1}(x_{2})\,\phi_\text{f}^{2n'}(x_{3})\>\,,
\end{align}
where $n'=n$, $n-1$ and $n'>1$.\footnote{For $n'=1$ and $n=2$, \eqref{boxk-1-1-2n} yields
\begin{align}
	2^{2k-1}(k-1)!\(1-2k\)_{k}\lim_{\e\rightarrow0}\(\a^{-1}\(2\g_{1}^{\phantom{()}\!\!}-\g_{2}^{\phantom{()}\!\!}\)\)=3\,,
\end{align}
which is consistent with \eqref{match-p} upon using \eqref{alpha-gamma}.
Moreover, as noticed in \cite{Nii:2016lpa}, the matching condition \eqref{boxk-boxk-1-1-2n} with $n'=1$ and $n=2$ is satisfied automatically given $\g_{2}^{\phantom{()}\!\!}=\frac{1}{3}\e+O(\e^{2})$, which is derived in \eqref{gamma-phi-p}.}
We can again use \eqref{3-point-spinning-simplified} and \eqref{Box-2p} to derive the independent constraints
\begin{align}\label{match-3}
4^{k}\left(\frac{(1-n')k}{n-1}\right)_{k}\left(\frac{(1-n'-n)k}{n-1}+1\right)_{k}
\lim_{\e\rightarrow 0}\left(\a^{-1}{\l_{\phi\phi\phi^{2n'}}}\right)
=\l_{2n-1,1,2n'}\,,
\end{align}
and
\begin{align}\label{match-4}
2^{2k-1}(k-1)!\left(1-\frac{n k}{n-1}\right)_{k}
\lim_{\e\rightarrow 0}\left(\a^{-1}\left(\g_1^{\phantom{()}\!\!}+\g_{2n-1}^{\phantom{()}\!\!}-\g_{2n}^{\phantom{()}\!\!}\right)\right)
&=\frac{\l_{2n-1,2n-1,2n}}
{\l_{1,2n-1,2n}}\,,
\\
\label{match-5}
2^{2k-1}(k-1)!\left(1-\frac{n k}{n-1}\right)_{k}
\lim_{\e\rightarrow 0}\left(\a^{-1}\left(-\g_1^{\phantom{()}}\!\!+\g_{2n-2}^{\phantom{()}\!\!}-\g_{2n-1}^{\phantom{()}\!\!}\right)\right)
&=\frac{\l_{2n-1,2n-2,2n-1}}
{\l_{1,2n-2,2n-1}}\,.
\end{align}
The solutions of \eqref{match-3} are presented in Appendix \ref{sec:b}. 
The constraint \eqref{match-4} can be viewed as \eqref{match-2} with $p=2n-1$. 
This also applies to \eqref{match-5}, which can be seen as the case with $p=2n-2$, 
as \eqref{match-1} implies 
\begin{align}\label{alpha-gamma}
\lim_{\e\rightarrow 0}\a^{-1}\g_1^{\phantom{()}}\!\!=0\,.
\end{align}
Therefore, the $p=2n-1,2n-2$ matching constraints can be written in a general $p$ form
\begin{align}\label{match-p}
	2^{2k-1}{(k-1)!}\left(1-\frac{n k}{n-1}\right)_{k}
	\lim_{\e\rightarrow 0}\left(\a^{-1}\left(\g_{p}^{\phantom{()}}\!\!-\g_{p+1}^{\phantom{()}}\right)\right)
	=\frac{\l_{2n-1,p,p+1}}
	{\l_{1,p,p+1}}\,.
\end{align}
This is related to the fact that $\phi^{2n-1}$ becomes a primary in the Gaussian limit, 
so the descendant corrections are of higher order in $\e$.\footnote{One needs to be more careful at subleading orders, as the differences between descendant and primary are not negligible. }
The case of $k=1,n=2$ was already noticed in \cite{Rychkov:2015naa}. 
A sum of \eqref{match-p} from $p=1$ to $p=2n-2$ gives
\begin{align}
	2^{2k-1}{(k-1)!}\left(1-\frac{n k}{n-1}\right)_{k}
	\lim_{\e\rightarrow 0}\left(\a^{-1}\left(\g_{1}^{\phantom{()}}\!\!-\g_{2n-1}^{\phantom{()}}\right)\right)
	=\sum_{p=1}^{2n-2}\frac{\l_{2n-1,p,p+1}}
	{\l_{1,p,p+1}}\,,
\end{align}  
where the ratio of three-point function coefficients is
\begin{align} 
\frac{\l_{2n-1,p,p+1}}
{\l_{1,p,p+1}}=\frac{p!(2n-1)!}{n!(n-1)!(p-n+1)!}\,.
 \end{align} 
Together with \eqref{gamma-descendant} and \eqref{alpha-gamma}, we obtain
\begin{align}
	\alpha=(-1)^{k+1}4^{k}k!&\frac{(n!)^{3}}{(2n)!(2n-1)!}\left(\frac{k}{n-1}+1\right)_{k-1}\epsilon+O(\e^2)\,.\label{Al}
\end{align}
Since the leading correction is of first order, we assume that the $\e$ expansion of the $\phi^{2n}$ theory 
gives rise to integer power series in $\e$. 
The matching constraint \eqref{match-1} gives
\begin{align}\label{gamma-1}
\g_1^{\phantom{()}\!\!}=2(-1)^{k+1}{n(n-1)}\[\frac{(n!)^2}{(2n)!}\]^3
\frac{\left(\frac k {n-1}+1\right)_{k-1}}{\left(\frac {nk} {n-1}\right)_k}\,\e^2+O(\e^3)\,,
\end{align}
where the first order anomalous dimension of $\phi$ vanishes $\g_1^{(1)}=0$. 
Now we can solve \eqref{match-p} and the solution reads
\begin{align}\label{gamma-phi-p}
	\gamma_{p}^{\phantom{()}\!\!}=\frac{n-1}{(n)_{n}}\,(p-n+1)_{n}\,\e+O(\e^2)\,,
\end{align}
which is independent of $k$. 
These general $k$ results for $\phi^p$ were obtained previously in \cite{Gliozzi:2016ysv,Gliozzi:2017hni}.\footnote{Note that \cite{Gliozzi:2016ysv,Gliozzi:2017hni} studied the recombination at the level of 
conformal block expansion of four-point functions, 
and considered $\phi^{2n-1}$ as a primary operator. } 
Using the result \eqref{gamma-phi-p}, we obtain
\begin{align}\label{irrelevant}
	\D_{\phi^{2n}}-d=(n-1)\e+O(\e^{2})\,,
\end{align}
which means that the operator $\phi^{2n}$ is irrelevant $\D_{\phi^{2n}}>d$.
Furthermore, we can use the matching condition \eqref{match-3} to determine the first-order terms of 
$\l_{\phi\phi\phi^{2(n-1)}},\l_{\phi\phi\phi^{2n}}$, which are given in Appendix \ref{sec:b}. 
Let us emphasize that the discussion of $\phi^p$ is based on the assumption that 
all $\phi^p$ with $p\neq 2n-1$ are primary operators.

Above, we use the matching condition associated with $\langle \phi\, \phi\, \mathcal O\rangle$ to 
determine the $\phi^{2n}$ and $\phi^{2(n-1)}$ anomalous dimensions.
It is more natural to consider the primary bilinear operators 
\begin{align}
\mathcal{J}^{(m)}_{\ell,\text{f}}\sim\phi_\text{f}\, \pa^\ell\Box^m\phi_\text{f}\,,\qquad (\ell=0,2,4,\ldots)\,,
\end{align} 
as they already appear in the free OPE $\phi_\text{f}\times\phi_\text{f}$. 
Although the would-be-eaten multiplets do not appear in this OPE due to their odd spin, 
it turns out that the action of Laplacians on $\phi$ still gives rise to useful matching constraints. 
As shown in \cite{Gliozzi:2017gzh}, one can indeed determine the leading anomalous dimensions of 
broken higher-spin currents for $k=1$ using the matching conditions. 
The discussions in \cite{Gliozzi:2017gzh} are based on five-point functions, but, in our opinion, 
the nontrivial constraints are encoded in two- and three-point functions 
if OPE associativity is not taken into account. 
Let us emphasize that we do not make use of the fact that the higher-spin currents are (partially) conserved in the free theories.
In fact, $\mathcal{J}^{(m)}_{\ell,\text{f}}$ are not (partially) conserved at low spin, i.e., at $\ell<c$, as indicated in \eqref{partial-conservation}.
In principle, the computation of anomalous dimensions can be generalized to other nondegenerate primaries, which do not need to obey any conservation law.\footnote{
In practice, there could exist some technical difficulties if one wants to generalize the computation to arbitrary nondegenerate primaries. 
First, we need to write down all the nondegenerate primaries, which requires some careful examination.
Second, we need to know the ratios of OPE coefficients in the matching conditions, 
which could be technically involved for arbitrarily complicated operators. 
In fact, the most technically challenging part of the present work is to obtain the ratios of OPE coefficients \eqref{RaON2}, \eqref{RaON1}. 
Third, another problem arises when the free correlator on the right-hand side of the matching condition vanishes.
This usually leads to the vanishing of the anomalous dimension at low order.
To compute the anomalous dimension at leading nonvanishing order, we need to know the expression of the first nonzero order on the right-hand side of the matching condition, which cannot be directly derived from the (generalized) free theory. }
See \cite{Gliozzi:2017gzh} for related discussions in the case of $k=1$.

To study the anomalous dimensions of broken currents, we consider the matching condition
\begin{align}\label{2BoxBi}
	\lim_{\e\rightarrow 0}\a^{-2}\langle\Box^{k}\phi(x_{1})\,\Box^{k}\phi(x_{2})\,\mathcal{J}^{(m)}_{\ell}(x_{3},z)\rangle
	=\langle\phi^{2n-1}_{\text{f}}(x_{1})\,\phi^{2n-1}_{\text{f}}(x_{2})\,\mathcal{J}^{(m)}_{\ell,\text{f}}(x_{3},z)\rangle\,.
\end{align}
We assume that $\mathcal{J}^{(m)}_{\ell}$ are primary operators at the interacting fixed points.
For the spinning case, the identity \eqref{Box-2p} has a simple generalization
\begin{align}
\Box_{x_1}\frac{\left(z\cdot x_{12}\right)^{\ell}}{x_{12}^{2\D+\ell}}
=4\left(\D+\frac \ell 2\right)\left(\D-\frac \ell 2-\frac d 2+1\right)\frac{\left(z\cdot x_{12}\right)^{\ell}}{x_{12}^{2(\D+1)+\ell}}\,.
\end{align}
Together with \eqref{3-point-spinning-simplified}, we obtain
\begin{align}\label{BoxFa}
	\lim_{\e\rightarrow 0}\[\a^{-2}4^{2k}
	\(\D_{\phi}-\frac 1 2\D_{\mathcal{J}^{(m)}_{\ell}}+\frac\ell 2\)_{2k}
	\left(\D_{\phi}-\frac 1 2\D_{\mathcal{J}^{(m)}_{\ell}}-\frac \ell 2-\frac d 2+1\right)_{2k}\]
	=R\,,
\end{align}
where $R$ is the ratio of three-point function coefficients\footnote{See Appendix \ref{sec:Ratio of 3 point function coefficients} for more details about the calculation of $R$.}
\begin{align}
\label{ratio-3pt-current}
R=\frac{\l_{\phi^{2n-1}_{\text{f}}\phi^{2n-1}_{\text{f}}\mathcal{J}^{(m)}_{\ell,\text{f}}}}{\l_{\phi_{\text{f}}^{}\,\phi_{\text{f}}^{}\,\mathcal{J}^{(m)}_{\ell,\text{f}}}}=(2n-1)^2(2n-2)!\,.
\end{align} 
According to \eqref{gamma-definition}, the anomalous dimension of $\mathcal{J}^{(m)}_{\ell}$ is defined by
\begin{align}
	\D_{\mathcal{J}^{(m)}_{\ell}}=2\(\frac{k}{n-1}-\frac{\e}{2}\)+2m+\ell+\g_{\mathcal{J}^{(m)}_{\ell}}.
\end{align}
The explicit matching constraint is
\begin{align}\label{BoxFa}
	\Scale[1]{ 
	\lim_{\e\rightarrow 0}\[\a^{-2}
	\(\g_{1}^{\phantom{()}\!\!}-\frac 1 2\g_{\mathcal{J}^{(m)}_{\ell}}\)
	\left(\g_{1}^{\phantom{()}\!\!}-\frac 1 2\g_{\mathcal{J}^{(m)}_{\ell}}+\frac{\e}{2}+1-\ell-m-\frac {d_\text{u}} 2\right)_{2k}\]
	=\frac{(2n-1)^2(2n-2)!}{4^{2k}(-m)_m (2k-m-1)!} }\,.
\end{align}
There are two possible scenarios: 
\begin{itemize}
\item
 $\g_{\mathcal J}^{\phantom{()}\!\!}\sim\e^2$
 
In the generic case, the remaining Pochhammer symbol $(\dots)_{2k}$ has a finite Gaussian limit, so we have
\begin{align}\label{gamma-spin-generic}
	\g_{\mathcal{J}^{(m)}_{\ell}}
	=2\g_{1}^{\phantom{()}\!\!}-2\a^{2}\,\frac{(2n-1)^2(2n-2)!}{4^{2k}(-m)_m (2k-m-1)!\left(m+\ell-k\frac{n-2}{n-1}\right)_{2k}}+O(\e^3)\,,
\end{align}
which is consistent with the double-twist asymptotic behavior $\g_{\mathcal J}\rightarrow 2\g_1^{\phantom{()}\!\!}$ at large spin $\ell$. 
Note that $\alpha$ is given in \eqref{Al} and $\g_{1}^{\phantom{()}\!\!}$ is given in \eqref{gamma-1}. 
As expected, the spin-2 operator on the highest trajectory $m=k-1$ has no anomalous dimension, 
i.e., $\g_{\mathcal J_{\ell=2}^{(m=k-1)}}=0$, since it corresponds to the stress tensor. 
The formula \eqref{gamma-spin-generic} generalizes the $k=1$ result in \cite{Gliozzi:2017gzh}.
\item
 $\g_{\mathcal J}^{\phantom{()}\!\!}\sim\e^1$
 
We should be more careful 
at low spin $\ell\leqslant 2k-m-d_\text{u}/2$ if the upper critical dimension $d_\text{u}=2nk/(n-1)$ is an even integer.
Since we assume that $k$ and $n-1$ have no common divisor, the upper critical dimension $d_{\text{u}}$ is an even integer if and only if $n=2$.
Then we have $d_\text{u}=4k$, so the exceptional operator is associated with $\ell=m=0$, 
i.e., the spin-$0$ operator on the lowest trajectory.
The matching constraint becomes
\begin{align}\label{matching-J-special}
	\lim_{\e\rightarrow 0}\frac{\g_{2}^{\phantom{()}\!\!}
	(\g_{2}^{\phantom{()}\!\!}-\e)}{\e^{2}}
	=-\frac{2}{9}\,,
\end{align}
where $\g_{1}$ is omitted and 
the $k$ dependence is canceled by $\a$ in \eqref{Al} with $n=2$. 
There are two solutions
\begin{align}\label{gamma-spin-special}
\g_{2}^{\phantom{()}\!\!}
=\frac{\e}{2}\pm\frac{\e}{6}+O(\e^2)\,,
\end{align}
which are related by the shadow transformation $\D\rightarrow d-\D$. 
For $n=2$ with general $k$, the $\g_{p}$ formula \eqref{gamma-phi-p} implies that the physical solution corresponds to the case with a minus sign.\footnote{The missing one can be viewed as a null state. 
See also Sec. 4.4.2 in \cite{Henriksson:2022rnm} 
for the perspectives from the mixing effects and the redundant operators \cite{Wegner:1976}. }

\end{itemize}

The formula for the higher-spin currents \eqref{gamma-spin-generic} with $n=2$ and $m=0$ is singular at $\ell=0$.
Motivated by analyticity in spin \cite{Caron-Huot:2017vep}, 
we perform the analytic continuation of the generic formula \eqref{gamma-spin-generic} with $n=2$ 
in the conformal spin $\bar h=\t/2+\ell$. 
Then we impose the matching condition on the spin-0 anomalous dimensions
\begin{align}
	\g_{2}^{\phantom{()}\!\!}=\left.\g_{\mathcal{J}}^{\phantom{()}\!\!}(\bar h)\right|_{\bar h\rightarrow k - \frac \e 2+\frac {\g_{2}^{\phantom{()}\!\!}}{2}}\,,
\end{align}  
which leads to the constraint on the anomalous dimension of $\phi^{2}$
\begin{align}
	\g_{2}^{\phantom{()}\!\!}=-\frac{2\e^{2}}{9} \frac{1}{\g_{2^{\phantom{()}}}\!\!\!\!-\e}+O(\e^{2})\,.
\end{align}
This is precisely the same quadratic equation as \eqref{matching-J-special}, 
so we again find the two solutions in \eqref{gamma-spin-special}. 
A first-order formula in $\e$ can be derived from a second-order one 
because the special cases are related to the poles in the general formula. 
The order of $\g_{\mathcal{J}}^{\phantom{()}\!\!}(\bar h)$ is lowered by one and becomes first order in $\e$ 
due to the factor $\g_{2}^{\phantom{()}\!\!}-\e$ in the denominator.  
The matching with the spin-0 data by analytic continuation was noticed in the analytic bootstrap study of the $k=1, n=2$ case \cite{Alday:2017zzv}. 
See also \cite{Henriksson:2018myn, Caron-Huot:2020ouj,Caron-Huot:2022eqs} for further investigations. 

\subsection{Analytic bootstrap}
\label{sec:Lorentzian Inversion}
In this subsection, we use the analytic bootstrap to examine if the multiplet recombination results are consistent with OPE associativity. 
More concretely, we deduce the implications of crossing constraints, i.e., spinning anomalous dimensions, 
from the Lorentzian inversion formula.    
Systematic investigations of the standard $\phi^4$ theory ($k=1, n=2$) 
had been carried out in \cite{Alday:2017zzv, Henriksson:2018myn}. 
We generalize the discussions of the leading anomalous dimensions to the $\Box^{k}$ theory with $\phi^{2n}$ interaction, based on the spin-0 input from the multiplet recombination method. 

We consider the four-point function of identical scalars
\begin{align}
	\<\phi(x_{1})\phi(x_{2})\phi(x_{3})\phi(x_{4})\>=\frac{\mathcal{G}(u,v)}{x_{12}^{2\Delta_{\phi}}x_{34}^{2\Delta_{\phi}}}\,,
\end{align}
where $u,v$ are the conformally invariant cross ratios
\begin{align}
	u=z\bar{z}=\frac{x_{12}^{2}x_{34}^{2}}{x_{13}^{2}x_{24}^{2}}\,,\qquad 
	v=(1-z)(1-\bar{z})=\frac{x_{14}^{2}x_{23}^{2}}{x_{13}^{2}x_{24}^{2}}\,.
\end{align}
The crossing equation reads
\begin{align}
	\mathcal{G}(u,v)=\frac{u^{\Delta_{\phi}}}{v^{\Delta_{\phi}}}\mathcal{G}(v,u)\,.
\end{align}
The conformal block expansion of the right-hand side reads
\begin{align}
	\mathcal{G}(v,u)=1+\sum_i \tilde\l_{\phi\phi\mathcal O_i}^2\,G_{\D_i,\ell_i}(1-z,1-\bar z)\,.
\end{align}
We have introduced the OPE coefficient $\tilde{\l}_{\phi\phi\mathcal O_i}=\l_{\phi\phi\tilde{\mathcal O_i}}$ where the two-point function of $\tilde {\mathcal O}$ is unit-normalized, i.e., $\l_{\tilde{\mathcal O}\tilde{\mathcal O}\mathds{1}}=1$. 
The explicit example of $\tilde \l^{2}_{\phi\phi\phi^{2(n-1)}}$ can be found in \eqref{lambda-tilde-lambda}. 
In the analytic bootstrap, 
the leading behavior is associated with the contribution of the identity operator, 
which implies the existence of double-twist trajectories \cite{Fitzpatrick:2012yx,Komargodski:2012ek}
\begin{align}
\t_{m,\ell}=2\D_\phi+2m+\dots\,,
\end{align}
with squared OPE coefficients \cite{Fitzpatrick:2011dm}
\begin{align}\label{GFF-OPE}
\tilde{\l}^2_{\phi\phi \mathcal O_{m,\ell}}=
\Scale[1.3]{
\frac{(1+(-1)^\ell)[(\D_\phi)_{m+\ell}(\D_\phi-d/2+1)_m]^2}
{m!\ell!(d/2+\ell)_m (2\D_\phi-d+1+m)_m(2\D_\phi+m+\ell-d/2)_m(2\D_\phi+2m+\ell-1)_\ell} }+\dots \,,
\end{align}
where the ellipses indicate subleading terms at large spin. 
For $m\geqslant k$, the order $\e^0$ term of the leading OPE coefficient vanishes 
due to the factor $(\D_\phi-d/2+1)_{m\geqslant k}\sim\g_1^{\phantom{()}\!\!}\sim\e^2$. 
The case of $m=k-1$ corresponds to the conserved currents, 
while those of $0\leqslant m < k-1$ are related to the partially conserved currents.  
According to the double-twist behavior, 
it is also natural to define the anomalous dimension $\tilde{\g}_{\mathcal{J}_{\ell}^{(m)}}$ 
with respect to $2\D_\phi$
\begin{align}
	\tilde{\g}_{\mathcal{J}_{\ell}^{(m)}}=\D_{\mathcal{J}_{\ell}^{(m)}}-(2\D_{\phi}+2m+\ell)\,,
\end{align}
which involves the full scaling dimension of $\D_\phi$. 

For simplicity, we focus on the case of the leading trajectory $\phi\pa^\ell\phi$ with $m=0$, 
but our analysis extends to higher twists.\footnote{Using the light cone expansion of conformal blocks in Appendix \ref{Lightcone expansion of conformal blocks}, we also examine the cases of subleading and sub-subleading trajectories.  
The details can be found in Appendix \ref{Subleading twist and sub-subleading twist}.} 
Accordingly, we take the light cone limit of the Lorentzian inversion formula, 
which reduces to the SL($2,\mathbb{R}$) inversion integral \cite{Alday:2017zzv,Caron-Huot:2017vep}
\begin{align}\label{SL2R-inversion}
	C(z,\bar h)=\frac{(2\bar h-1)\Gamma(\bar{h})^{4}}{\pi^{2}\Gamma(2\bar{h})^2}\int_{0}^{1}\frac{\mathrm{d}\bar{z}}{\bar{z}^{2}}\bar{z}^{\bar{h}}{}_{2}F_{1}(\bar{h},\bar{h},2\bar{h},\bar{z})\,\text{dDisc}[\mathcal{G}(u,v)]\Big|_{z\rightarrow 0}\,,
\end{align}
where the double discontinuity is defined by analytic continuation around $\bar{z}=1$
\begin{align}
	\text{dDisc}[f(\bar{z})]=f(\bar{z})-\frac{1}{2}f^{\circlearrowleft}(\bar{z})-\frac{1}{2}f^{\circlearrowright}(\bar{z})\,.
\end{align}
To derive the anomalous dimensions of the leading trajectory, 
we expand the inversion result as
\begin{align}
C(z,\bar h)=z^{\D_\phi}\,\tilde{\l}_{\phi\phi\mathcal O}^2 \,
\Big(1+\frac {1}{2}\tilde\g_{\mathcal O}\log z+\dots\Big)+\dots\,,
\end{align}
so the leading anomalous dimension can be obtained by 
dividing the coefficients of the $\log z$ term by \eqref{GFF-OPE}. 
The concrete inversion procedure depends on $n$, 
which can be divided into two types:
\begin{itemize}
	\item
	Type I: $n=2$
	
	This is the $k>1$ generalization of the standard $\phi^4$ theory at $d=4-\e$. 
	\item
	Type II: $n>2$ ($k$ and $n-1$ have no common divisor)
	
	This generalizes the standard $\phi^6$ theory at $d=3-\e$ to generic $k, n$. 
\end{itemize}
In both cases, the leading corrections are associated with the cross-channel scalar $\phi^{2(n-1)}$ with 
\begin{align}
\D_{\phi^{2(n-1)}}=2k+O(\e)\,.
\end{align} 
For Type I, we have $\phi^{2(n-1)}=\phi^2$, which appears already in the free OPE $\phi_\text{f}\times \phi_\text{f}$, 
and the leading contribution is associated with squared anomalous dimension $(\g_{2})^2\sim \e^2$. 
For Type II, the free OPE $\phi_\text{f}\times \phi_\text{f}$ does not contain $\phi^{2(n-1)}_\text{f}$, 
but this operator can appear in the $\phi\times\phi$ OPE because the $\phi^{2n}$ interaction leads to a first order OPE coefficient, 
i.e., $\tilde\l_{\phi\phi\phi^{2(n-1)}}^2\sim\e^2$. 
To derive the anomalous dimensions, 
we need to know the $\log z$ term of the light cone expansion of the $\phi^{2(n-1)}$ block
\begin{align}
G^{(d)}_{\D,0}(1-\bar{z},1-z)\big|_{\log z}
=(-1)\frac{\Gamma(\D)}{\Gamma(\D/2)^{2}}(1-\bar{z})^{\D/2}\,_{2}F_{1}
\(\frac{\D}{2},\frac{\D}{2},\D-\frac{d}{2}+1,1-\bar{z}\)
+O(z)\,,\quad
\end{align}
which should be multiplied by $z^{\D_\phi}\zb^{\D_\phi}/(1-\zb)^{\D_\phi}$ before evaluating the double discontinuities. 

\subsubsection*{Type I: $n=2$}
Let us explain why the leading corrections of anomalous dimensions are associated with $\phi^2$ in the $\phi^4$ theory with $k\geqslant 1$. 
Since $\D_{\phi_\text{f}}=k$, the double discontinuity of a $\mathbb Z_2$-even operator $\pa^{2i_1}\phi^{2 i_2}$ in the $\phi\times\phi$ OPE is proportional to $\tilde\g_{\mathcal O}^2$. 
To compute second order corrections in $\e$, we need to use $\tilde{\l}_{\phi\phi\mathcal O}\sim \e^0$ and $\tilde\g_{\mathcal O}\sim\e$.\footnote{We assume that the $\e$ expansion of the CFT data gives integer power series in $\e$.} 
The first condition restricts the choice to the double-twist operators $\phi\,\pa^\ell\Box^m\phi$ with $m\leqslant k-1$. 
For the second condition, the lowest scalar $\phi^2$ is special in that it is the only one with nonzero anomalous dimension $\tilde\g$ 
at order $\e$.  
The leading double discontinuity of $\mathcal{G}(u,v)$ is given by
\begin{align}
	-\frac{\p^2}{2}\frac{\Gamma(\D)}{\Gamma(\D/2)^{2}}\tilde \l_{\phi\phi\phi^{2}}^{2}(\tilde\g_{\phi^{2}})^{2}\,
	z^{\D_{\phi}}\zb^{\D_{\phi}}\,_{2}F_{1}\(\frac{\D_{\phi^{2}}}{2},\frac{\D_{\phi^{2}}}{2},\D_{\phi^{2}}-\frac{d}{2}+1,1-\bar{z}\)\,,
\end{align}
where $\text{dDisc}[\log^{2}(1-\zb)]=4\p^{2}$ is used. 
According to \eqref{gamma-phi-p}, the first-order anomalous dimension of $\phi^2$ is
\begin{align}\label{phi2-gamma}
\tilde\g_{\phi^2}^{(1)}=\g_{2}^{(1)}=\frac 1 3\,,
\end{align}
so it is independent of $k$. 
The free theory values for the other input parameters are
\begin{align}
	\tilde \l_{\phi\phi\phi^{2}}^{2}=2+O(\e),\quad\D_{\phi^{2}}=2k+O(\e),\quad d=4k+O(\e)\,.
\end{align}
Then we evaluate the inversion integral \eqref{SL2R-inversion} and perform the substitution 
$\bar h\rightarrow k+\ell+O(\e)$. 
The result reads
\begin{align}\label{tildegamma-n=2}
\tilde\g_{\mathcal{J}_{\ell}^{(0)}}=-\frac{(2k-1)!(\ell-1)!}{9(2k+\ell-1)!}\e^{2}+O(\e^{3})\,,
\end{align}
which is precisely \eqref{gamma-spin-generic} with $n=2, m=0$. 
We also confirm the consistency for $m=1,2$ 
and more details can be found in appendix \ref{Subleading twist and sub-subleading twist}.

\subsubsection*{Type II: $n>2$ ($k$ and $n-1$ have no common divisor)}
\begin{figure}[h]
	\centering
	\includegraphics[width=.55\textwidth,origin=c,angle=0]{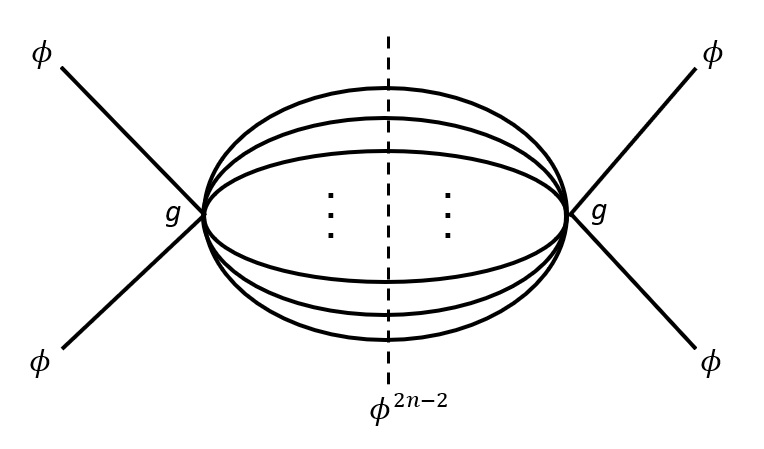}
	\caption{In the Lagrangian description of a type II theory, 
	the leading correction from $\phi^{2n-2}$ contributes at order $g^{2}\sim\e^2$ due to the $g\phi^{2n}$ vertices. \label{Type2}}
\end{figure}

As the anomalous dimensions of all double-twist operators are of second order in $\e$, 
their double discontinuities are of order $\e^4$. 
We should instead consider new operators that are absent in the free OPE $\phi_\text{f}\times\phi_\text{f}$. 
As discussed in \cite{Henriksson:2020jwk}, a natural candidate in $\phi^{2n}$ theory is the scalar $\phi^{2(n-1)}$ (see Fig. \ref{Type2}). 
The OPE coefficient can also be deduced from the multiplet recombination \eqref{match-3}
\begin{align}\label{lambda-tilde-lambda}
	\tilde \l^{2}_{\phi\phi\phi^{2(n-1)}}=\frac{\l^{2}_{\phi\phi\phi^{2(n-1)}}}{(2n-2)!}\,,
\end{align}
where the leading term of $\l_{\phi\phi\phi^{2n-2}}$ can be found in Appendix \ref{sec:b}. 
Since $\D_{\phi^{2(n-1)}}=2k+O(\e)$, the leading double discontinuity reads
\begin{align}
-2\sin\(\frac{\p(n-2)k}{n-1}\)\tilde \l^{2}_{\phi\phi\phi^{2n-2}}z^{\Delta_{\phi}}\zb^{\frac{k}{n-1}}(1-\zb)^{\frac{(n-2)k}{n-1}}\frac{\Gamma(2k)}{\Gamma(k)^{2}}\,_{2}F_{1}\!\(k,k,2k-\frac{nk}{n-1}+1,1-\zb\).
\end{align}
Then we evaluate the inversion integral \eqref{SL2R-inversion} and make the substitution 
$\bar h\rightarrow k/(n-1)+\ell+O(\e)$. 
The result again agrees with the generic formula \eqref{gamma-spin-generic} from the multiplet recombination. 
We also verify the consistency between the results from multiplet recombination and those from analytic bootstrap for the subleading and sub-subleading trajectories.

The explicit expression of $\l_{\phi\phi\phi^{2n-2}}$ in Appendix \ref{sec:b} is not valid for $n=2$,  
as it diverges in the $n\rightarrow2$ limit due to the assumption of vanishing zeroth-order term. 
However, since the double discontinuity goes to zero,
the formal product of $\text{dDisc}[(1-\zb)^{\frac{(n-2)k}{n-1}}]$ and 
$\tilde{\l}_{\phi\phi\phi^{2n-2}}^2$ remains finite in the singular limit $n\rightarrow2$. 
The resulting inversion integral yields the same result as \eqref{tildegamma-n=2}. 
This matching from analytic continuation in $n$ is similar in spirit to 
the analytic continuation in spin of the second-order formula \eqref{gamma-spin-generic}, 
which reproduces the first-order results in the singular spin-0 limit.

\section{O($N$) models}
\label{sec:O(N) models}
We would like to generalize the results in Sec. \ref{sec:Anomalous dimensions of broken partially conserved currents} 
to the cases with global O($N$) symmetry. 
We use $\varphi_a$ to denote the fundamental field transforming in the vector representation. 
In the free theory, the higher-spin currents are bilinear operators of the schematic form 
$\varphi\,\pa^\ell\,\Box^m\varphi$.  
The tensor product decomposition reads:
\begin{align}
	\text{V}\otimes \text{V}=\text{S}\oplus \text{T}\oplus \text{A}
	=\bullet\oplus\ydiagram{2}\oplus\ydiagram{1,1}\,,
\end{align}
where V is the vector representation, S is the singlet representation, T is the rank-$2$ symmetric traceless representation and A is the rank-2 anti-symmetric representation. 
As a result, we have three sets of broken currents with different anomalous dimensions. 
We refer to \cite{Henriksson:2022rnm} for a recent review about the $\varphi^4$ O($N$) model with a canonical kinetic term. 

\subsection{Multiplet recombination}
The $N=1$ recombination equation \eqref{rec} 
has a direct generalization for general $N$:
\begin{align}\label{recombination-ON}
	\lim_{\e\rightarrow0}\alpha_N^{-1}\Box^{k}\varphi_{a}=\left(\varphi_{a}\varphi^{2n-2}\right)_\text{f}\,,
\end{align}
where $\varphi^{2q}\equiv(\sum_{a=1}^N\varphi_a \varphi_a)^{q}$ and we use the subscript $N$ to indicate the O($N$) generalization. 
To remind the reader, we use $\text{f}$ to indicate free theory values 
and we will write it as a subscript of $\langle\ldots\rangle$ to simplify the notation. 
The matching condition of the two-point function 
\begin{align}
	\lim_{\e\rightarrow0}\alpha_N^{-2}\<\Box^{k}\varphi_a(x_{1})\,\Box^{k}\varphi_b(x_{2})\>&
	=\<\varphi_{a}\varphi^{2n-2}(x_{1})\varphi_{b}\varphi^{2n-2}(x_{2})\>_\text{f}
\end{align}
gives 
\begin{align}\label{GaAlON}
	4^{2k}\(\frac{k}{n-1}\)_{2k}(1-k)_{k-1}k!\,\lim_{\e\rightarrow 0}\left(\a^{-2}\g_\varphi\right)
	=4^{n-1}(n-1)!\(1+\frac{N}{2}\)_{n-1}\,.
\end{align}
The O($N$) generalizations of the three-point matching condition \eqref{Box3P} are:
\begin{align}
	\label{match-ON-1}
	\lim_{\e\rightarrow 0}\a_N^{-1}\<\Box^{k}\varphi_{a}(x_{1})\varphi_{b}\varphi^{2q-2}(x_{2})\varphi^{2q}(x_{3})\>&=\<\varphi_{a}\varphi^{2n-2}(x_{1})\varphi_{b}\varphi^{2q-2}(x_{2})\varphi^{2q}(x_{3})\>_\text{f}\,,\\
	\label{match-ON-2}
	\lim_{\e\rightarrow 0}\a_N^{-1}\<\Box^{k}\varphi_{a}(x_{1})\varphi^{2q}(x_{2})\varphi_{b}\varphi^{2q}(x_{3})\>&=\<\varphi_{a}\varphi^{2n-2}(x_{1})\varphi^{2q}(x_{2})\varphi_{b}\varphi^{2q}(x_{3})\>_\text{f}\,.
\end{align}
As the recursion relations involve operators in the singlet and vector representations, there are two scenarios corresponding to \eqref{match-ON-1} and \eqref{match-ON-2}.
The explicit constraints are again given by \eqref{match-p} 
\begin{align}
	\label{recursion-ON-1}
	2^{2k-1}{(k-1)!}\left(1-\frac{n k}{n-1}\right)_{k}
	\lim_{\e\rightarrow 0}\left(\a^{-1}_{N}\left(\gamma_{2q-1}^{\phantom{()}\!\!}-\gamma_{2q}^{\phantom{()}\!\!}\right)\right)
	&=\frac{\l_{2n-1,2q-1,2q}}{\l_{1,2q-1,2q}}\,,
	\\
	\label{recursion-ON-2}
	2^{2k-1}{(k-1)!}\left(1-\frac{n k}{n-1}\right)_{k}
	\lim_{\e\rightarrow 0}\left(\a^{-1}_{N}\left(\gamma_{2q}^{\phantom{()}\!\!}-\gamma_{2q+1}^{\phantom{()}\!\!}\right)\right)
	&=\frac{\l_{2n-1,2q,2q+1}}{\l_{1,2q,2q+1}}\,,
\end{align}
where $\phi^p$ is replaced by $\varphi^{2q}$ for even $p$ and substituted by 
$\{\varphi_{a}\varphi^{2q}, \varphi_{b}\varphi^{2q}\}$ for odd $p$. 
Here $\g_{2q}^{\phantom{()}\!\!}$ is the anomalous dimension of $\varphi^{2q}$, and $\g_{2q+1}^{\phantom{()}\!\!}$ is the anomalous dimension of $\varphi_{a}\varphi^{2q}$
\begin{align}
	\g_{2q}^{\phantom{()}\!\!}\equiv\g_{\varphi^{2q}}\,,\qquad \g_{2q+1}^{\phantom{()}\!\!}\equiv\g_{\varphi^{2q+1}}\,,
\end{align}
where $\varphi^{2q+1}$ means $\varphi_{a}\varphi^{2q}$ with the O($N$) index suppressed.
The free three-point function coefficient of $\varphi^{q_1}_{\text{f}}$, $\varphi^{q_2}_{\text{f}}$, and $\varphi^{q_3}_{\text{f}}$ is denoted by $\l_{q_1,q_2,q_3}$
\begin{align}
	\l_{q_1,q_2,q_3}\equiv\l_{\varphi^{q_1}_{\text{f}}\varphi^{q_2}_{\text{f}}\varphi^{q_3}_{\text{f}}}\,.
\end{align}
The ratios of three-point function coefficients for general $q, N$ are\footnote{Let us explain how we derive these ratios. 
The coefficient of $N^i$ in the $N$ expansion can be computed by counting the number of configurations with $i$ loops. 
Based on some concrete examples, 
we notice that the decomposition of these ratios in terms of $(1-n-N/2)_j$ leads to simple coefficients 
in which the $q$ dependence is encoded in Pochhammer symbols. 
Then we use the $N=1$ expression to fix the complete coefficients and determine the general $N$ formulas.}

\begin{align}
	\frac{\l_{2n-1,2q-1,2q}}{\l_{1,2q-1,2q}}&=\frac{2^{n-2}(n+1)_{n}(q-n+1)_{n-1}}{n!}\,_{3}F_{2}\[\begin{matrix}\frac{1}{2}-\frac{n}{2},-\frac{n}{2},1-n-\frac{N}{2}\\ q-n+1,\frac{1}{2}-n\end{matrix};1\]\,,\label{RaON2}\\
	\frac{\l_{2n-1,2q,2q+1}}{\l_{1,2q,2q+1}}&=\frac{2^{n-2}(n+1)_{n}(q-n+2)_{n-1}}{n!}\,_{3}F_{2}\[\begin{matrix}\frac{1}{2}-\frac{n}{2},1-\frac{n}{2},1-n-\frac{N}{2}\\ q-n+2,\frac{1}{2}-n\end{matrix};1\]\,.\label{RaON1}
\end{align}
The ${}_{3}F_{2}$ hypergeometric series terminate for integer $n$ due to the first two parameters. 
A sum of \eqref{recursion-ON-1} and \eqref{recursion-ON-2} from $q=1$ to $q=n-1$ gives
\begin{align}
	\Scale[0.95]{
	2^{2k-1}{(k-1)!}\left(1-\frac{n k}{n-1}\right)_{k} }
	\lim_{\e\rightarrow 0}\left(\a^{-1}_{N}\left(\gamma_{1}^{\phantom{()}\!\!}-\gamma_{2n-1}^{\phantom{()}\!\!}\right)\right) 
	=\sum_{q=1}^{n-1} 	 \(\frac{\l_{2n-1,2q-1,2q}}{\l_{1,2q-1,2q}}+\frac{\l_{2n-1,2q,2q+1}}{\l_{1,2q,2q+1}}\)\,.
\end{align}
As in the $N=1$ case, we have $\lim_{\e\rightarrow 0}(\a^{-1}_{N}\,\g_1^{\phantom{()}\!\!})=0$ and $\lim_{\e\rightarrow 0}(\a^{-1}_{N}\,\g_{2n-1}^{\phantom{()}\!\!})=(n-1)\lim_{\e\rightarrow0}(\a_{N}^{-1}\,\e)$ due to \eqref{recombination-ON} and \eqref{GaAlON}. We obtain
\begin{align}\label{AlON}
\a_N=\frac{(-1)^{k+1}2^{2k-n}(n-1)(k-1)!\(\frac{k}{n-1}+1\)_{k}}{(n)_{n}\,_{3}F_{2}\[\begin{matrix}\frac{1}{2}-\frac{n}{2},-\frac{n}{2},1-n-\frac{N}{2}\\ 1,\frac{1}{2}-n\end{matrix};1\]}\,\e+O(\e^2)\,,
\end{align}
where the ${}_3F_2$ hypergeometric series is associated with summing those appearing in the ratios of three-point function coefficients. 
Then the solution to \eqref{GaAlON} reads
\begin{align}\label{gamma-1-ON}
\gamma_{1}^{\phantom{()}\!\!}=\frac{4^{n-2k-1}(n-1)!\(1+\frac{N}{2}\)_{n-1}}{\(\frac{k}{n-1}\)_{2k}(1-k)_{k-1}k!}\a^2_{N}+O(\e^3)\,,
\end{align}
which is the $k>1$ generalization of the eq. (4.103) in \cite{Wegner:1976}.
As before, the first order anomalous dimension of $\varphi_{a}$ vanishes, i.e., $\g_{1}^{(1)}=0$.
The anomalous dimensions for other scalar primaries are
\begin{align}\label{gamma-2q-V}
	\gamma_{2q+1}^{\phantom{()}\!\!}=\frac{2q-n+2}{n(n-2)!}\(q-n+2\)_{n-1}\frac{\,_{3}F_{2}\[\begin{matrix}\frac{1-n}{2},-\frac{n}{2},1-n-\frac{N}{2}\\ q-n+2,\frac{1}{2}-n\end{matrix};1\]}{\,_{3}F_{2}\[\begin{matrix}\frac{1-n}{2},-\frac{n}{2},1-n-\frac{N}{2}\\ 1,\frac{1}{2}-n\end{matrix};1\]}+O(\e^2)\,,
\end{align}
and
\begin{align}
\label{ON-phi-2a}
	\gamma_{2q}^{\phantom{()}\!\!}=\,&\gamma_{2q+1}^{\phantom{()}\!\!}-\frac{\(q-n+2\)_{n-1}}{(n-2)!}\frac{{}_{3}F_{2}\[\begin{matrix}\frac{1-n}{2},1-\frac{n}{2},1-n-\frac{N}{2}\\ q-n+2,\frac{1}{2}-n\end{matrix};1\]}{{}_{3}F_{2}\[\begin{matrix}\frac{1-n}{2},-\frac{n}{2},1-n-\frac{N}{2}\\ 1,\frac{1}{2}-n\end{matrix};1\]}+O(\e^2)\,.
\end{align}
These generalize the $n=2,3$ results obtained in \cite{Gliozzi:2016ysv,Gliozzi:2017hni}.
Moreover, we obtain an O($N$) generalization of \eqref{irrelevant}
\begin{align}
	\D_{\varphi^{2n}}-d=(n-1)\e+O(\e^{2})\,,
\end{align}
which is independent of $N$.
Therefore, $\varphi^{2n}$ is an irrelevant operator with $\D_{\varphi^{2n}}>d$.

Then we consider the matching conditions involving the broken higher-spin currents
\begin{align}
	\lim_{\e\rightarrow 0}\alpha_N^{-2}\<\Box^{k}\varphi_{a}(x_{1})\Box^{k}\varphi_{b}(x_{2})\mathcal{J}^{(m)}_{\ell}(x_{3},z)\>=\<\varphi_{a}\varphi^{2n-2}(x_{1})\varphi_{b}\varphi^{2n-2}(x_{2})\mathcal{J}^{(m)}_{\ell}(x_{3},z)\>_\text{f}\,.\qquad
\end{align}
The explicit constraints can be derived from \eqref{BoxFa} 
by substituting $\phi^p$ with  $\varphi^{2q}$ for even $p$ and 
$\{\varphi_{a}\varphi^{2q}, \varphi_{b}\varphi^{2q}\}$ for odd $p$
\begin{align}\label{matchingON}
	\lim_{\e\rightarrow 0} 	\Scale[0.95]{
	\[\a^{-2}_{N}
	\(\g_{1}^{\phantom{()}\!\!}-\frac 1 2\g_{\mathcal{J}^{(m)}_{\ell}}\)
	\left(\g_{1}^{\phantom{()}\!\!}-\frac 1 2\g_{\mathcal{J}^{(m)}_{\ell}}+\frac{\e}{2}+1-\ell-m-\frac {d_\text{u}} 2\right)_{2k}\] }
	=\frac{R_{N}}{4^{2k}(-m)_m (2k-m-1)!}\,.
\end{align}
The solutions of  three kinds of broken currents depend on the corresponding ratios of three-point function coefficients
\begin{align}\label{RaBiON}
	R_N=\frac{\l_{\varphi^{2n-1}\varphi^{2n-1}\mathcal{J}_{\ell}^{(m)}}}{\l_{\varphi\varphi\mathcal{J}_{\ell}^{(m)}}}=
	\begin{cases}
		4^{n-1}(2n-1)(n-1)!\left(\frac{N}{2}+1\right)_{n-1}&\quad\!\!\!\!\!\text{singlet}\\
		2^{2n-3}(4n+N-2)(n-1)!\left(\frac{N}{2}+2\right)_{n-2}&\quad\!\!\!\!\!\text{symmetric traceless}\\
		4^{n-1}(n-1)!\left(\frac{N}{2}+1\right)_{n-1}&\quad\!\!\!\!\!\text{anti-symmetric,}
	\end{cases}
\end{align}
which are derived in Appendix \ref{sec:Ratio of 3 point function coefficients}. 
As in the $N=1$ case, we find two possibilities:
\begin{itemize}
\item
 $\g_{\mathcal J}\sim\e^2$

In the generic case, we have
\begin{align}\label{BiON1}
	\gamma_{\mathcal{J}^{(m)}_{\ell}}=2\g_{1}^{\phantom{()}\!\!}-2\a_N^{2}\frac{R_N}{4^{2k}(-m)_m(2k-m-1)!\left(m+\ell-k\frac{n-2}{n-1}\right)_{2k}}+O(\e^3)\,,
\end{align}
which takes almost the same form as \eqref{gamma-spin-generic}. 
The anomalous dimensions of the broken currents are expressed in terms of $\alpha_{N}$ in \eqref{AlON}, $\g_{1}^{\phantom{()}\!\!}$ in \eqref{gamma-1-ON} and $R_{N}$ in \eqref{RaBiON}. 
%Note that $\a_N$ is given in \eqref{AlON}. 
One can check that the spin-2 singlet and the spin-1 antisymmetric operators on the highest trajectories, i.e.,  $m=k-1$, 
have vanishing anomalous dimensions. 
They correspond to the stress tensor and the O($N$) symmetry currents. 

\item
 $\g_{\mathcal J}\sim\e^1$
 
As in the $N=1$ case, we should be more careful when $\ell\leqslant 2k-m-d_\text{u}/2$ with even $d_\text{u}$.
The assumption that $k$ and $n-1$ have no common divisor implies $n=2$ at even $d_{\text{u}}$.
Again, the exceptional operators are the spin-$0$ operators on the lowest trajectory with $m=0$.
The matching condition \eqref{matchingON} then gives
\begin{align}\label{BiON2}
\g_{\varphi^{2}_{\text{S}}}
=\frac{\e}{2}\pm\frac{4-N}{2(N+8)}\e+O(\e^2)\,,\qquad
\g_{\varphi^{2}_{\text{T}}}
=\frac{\e}{2}\pm\frac{N+4}{2(N+8)}\e+O(\e^2)\,,
\end{align}
which are independent of $k$ at leading order. 
The solutions with different signs are related by the shadow transform $\D\rightarrow d-\D$. 
Note that $\varphi_{\text{S}}^{2}=\varphi^{2}$ is the singlet operator, and $\varphi_{\text{T},ab}^{2}=\varphi_a\varphi_b-\frac{\d_{ab}}{N}\varphi^{2}$ is the symmetric traceless operator.
The spin-$0$ operator on the lowest trajectory in the anti-symmetric representation does not exist. 
The physical solutions correspond to those with the minus sign. 
The case of $\g_{\varphi^{2}_{\text{S}}}$ can be obtained from $\g_{2q}^{\phantom{()}\!\!}$ formula \eqref{ON-phi-2a}.\footnote{\label{phi2-T}
As in \cite{Rychkov:2015naa}, the anomalous dimension of the symmetric traceless operator $\varphi^{2q}_{\text{T}}$ can be derived from the matching condition with $n=2$
\begin{align}
	\lim_{\e\rightarrow 0}\a_N^{-1}\<\Box^{k}\varphi_{a}(x_{1})\varphi_{b}\varphi^{2q-2}(x_{2})\varphi^{2q}_{\text{T,ce}}(x_{3})\>=\<\varphi_{a}\varphi^{2}(x_{1})\varphi_{b}\varphi^{2q-2}(x_{2})\varphi^{2q}_{\text{T,ce}}(x_{3})\>_\text{f}\,.
\end{align}
The matching condition above leads to
\begin{align}
	2^{2k-1}(k-1)!\(1-2k\)_{k}\lim_{\e\rightarrow0}\a^{-1}\(\g_{\varphi^{2q-1}}-\gamma_{\varphi_{\text{T}}^{2q}}\)=2(3p-2)\,.
\end{align}
From the solutions of $\a$ and $\g_{\varphi^{2q-1}}$ in \eqref{AlON} and \eqref{gamma-2q-V}, we obtain
\begin{align}\label{ST-gamma}
	\gamma_{\varphi_{\text{T}}^{2q}}=\frac{-N+q(N+6q-4)}{N+8}\,,
\end{align}
which is independent of $k$. 
In particular, the $q=1$ case gives $\gamma_{\varphi^{2}_{\text{T}}}=\frac{2}{N+8}\epsilon+O(\epsilon^{2})$.
In the $q=2$ case, we verify that $\varphi_{\text{T}}^{4}$ is an irrelevant operator because $\D_{\varphi_{\text{T}}^{4}}-d=\frac{8}{N+8}\e+O(\e^{2})$.
See also \cite{Gliozzi:2016ysv,Gliozzi:2017hni} for the case of $n=3$. } 

\end{itemize}
As in the $N=1$ case \cite{Alday:2017zzv}, we can also obtain the spin-$0$ anomalous dimensions \eqref{BiON2} from \eqref{BiON1} by analytic continuation in conformal spin \cite{Henriksson:2018myn}. 
According to \eqref{BiON2}, the two solutions in the singlet case are degenerate at $N=4+O(\e)$. 
To have a better understanding of the $N$ dependence, 
we examine the Chew-Frautschi plot at different $N$. 
%We consider the trajectories of singlet bilinear operators with minimal twists in $\varphi^{4}$ theory.
Following \cite{Caron-Huot:2022eqs}, the analytic continuation of the leading trajectories 
with lowest twist, i.e., $\t=2\D_\varphi+O(\e^2)$, is given by
\begin{align}\label{trajectory}
(\D-d/2)^2&=\(2k-\e+\ell+\gamma_{\mathcal{J}^{(0)}_{\ell}}-d/2\)^{2}\nn
&=\ell^2-\ell\e+\(\frac{1}{4}+\(\frac{(-1)^{k+1}(k+1)_{k-1}}{(2k)_k}\ell-\frac{3(2k-1)!}{(\ell+1)_{2k-1}}\)\frac{2(N+2)}{(N+8)^2}\)\e^2+O(\e^3)\,,
\end{align}
where the $1/\ell$ poles cancel at each order in the $\e$ expansion and $(\D-d/2)$ is analytic in $\ell$ near $\ell=0$. 
Setting $\ell=0$, we obtain the same solutions \eqref{BiON2} from the analytic continuation in conformal spin. 

\begin{figure}[h]
	\centering
	\includegraphics[width=.7\textwidth,origin=c,angle=0]{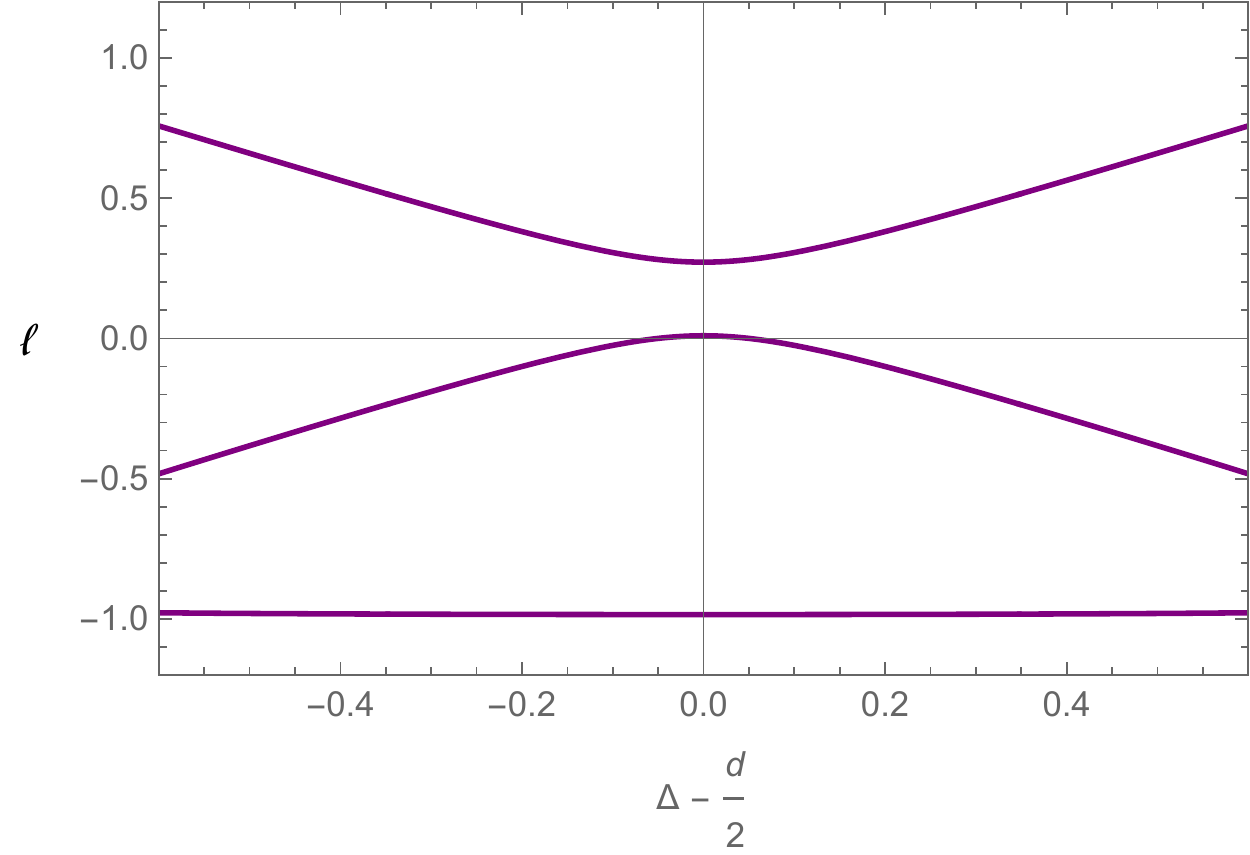}
	\caption{Chew-Frautschi plot of the leading Regge trajectory near the intercept from \eqref{trajectory} with $k=1,N=1,\e=0.3$.}
	\label{CF-plot}
\end{figure}

\begin{figure}[h]
	\centering
	\begin{subfigure}{.5\textwidth}
		\raggedright
		\includegraphics[width=.9\linewidth]{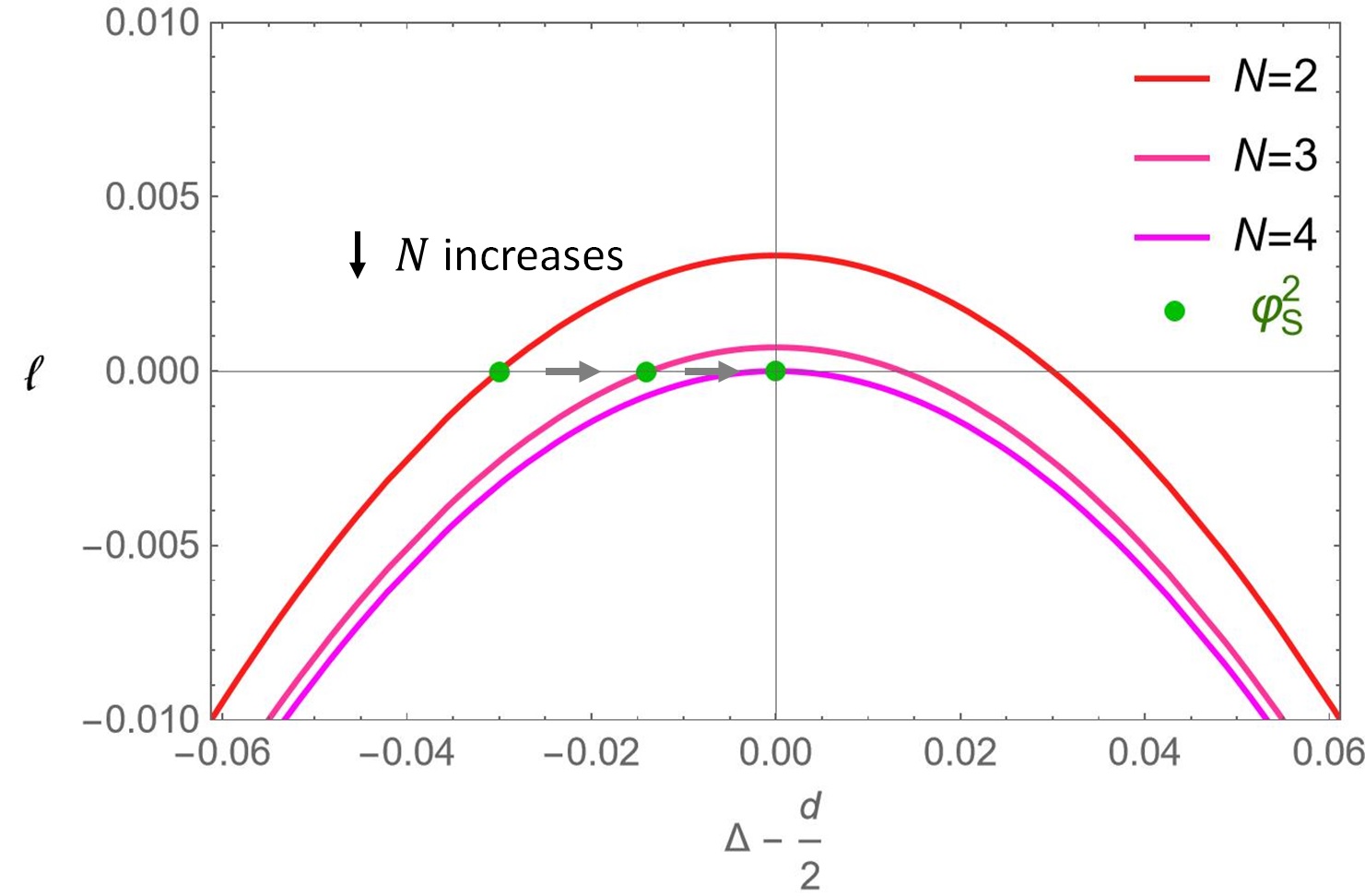}
		\caption{$N\leqslant4$}
	\end{subfigure}%
	\begin{subfigure}{.5\textwidth}
		\raggedright
		\includegraphics[width=.9\linewidth]{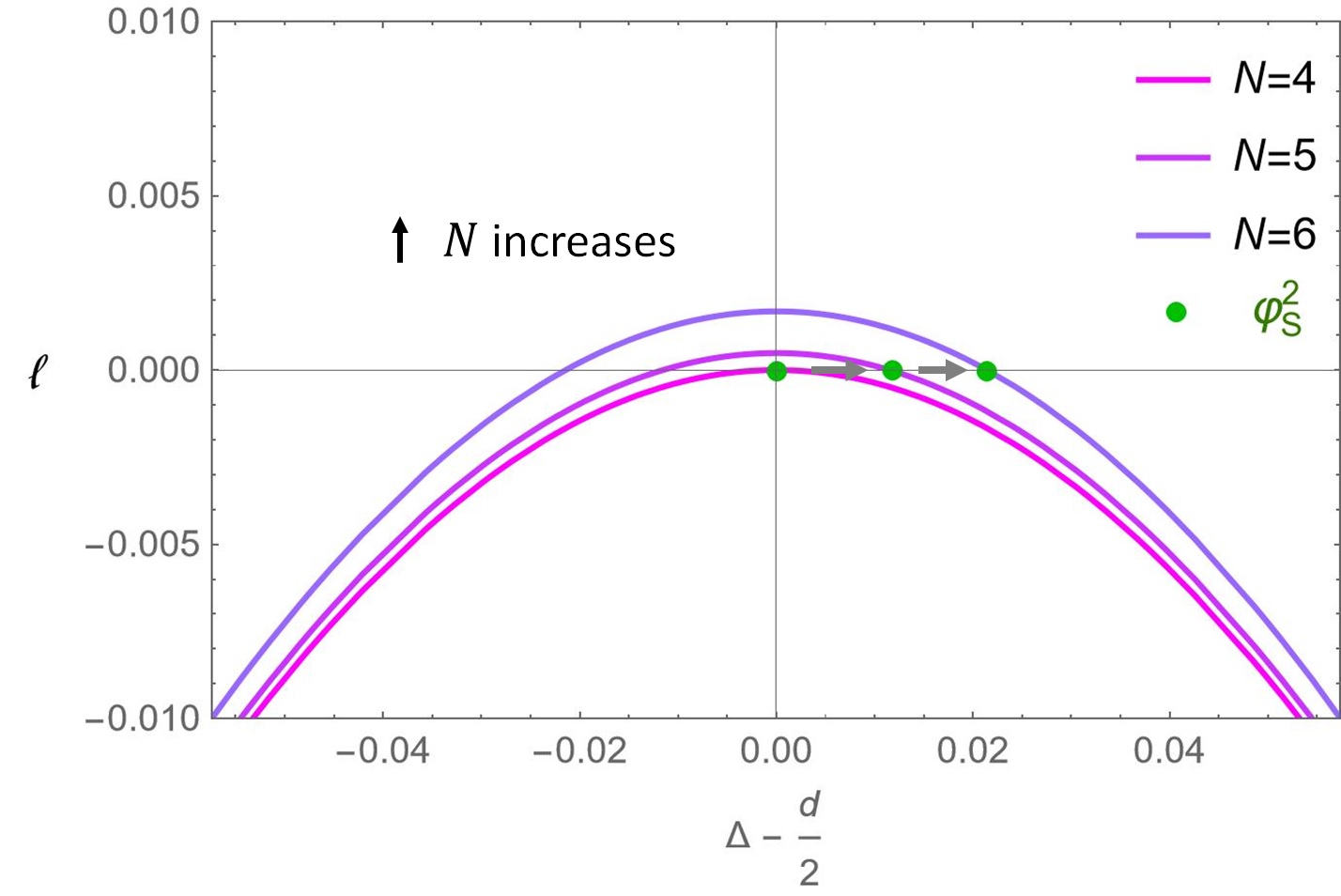}
		\caption{$N\geqslant4$}
	\end{subfigure}
	\caption{The singlet trajectories near $\ell=0$ from \eqref{trajectory}. The intersection associated with the singlet operator $\varphi_{\text{S}}^{2}$ is represented by the green points. As $N$ increases, the trajectory near $\ell=0$ moves downward for $N<4$ and then upward for $N>4$. 
The physical intersection associated with $\varphi_{\text{S}}^{2}$ moves from left to right. At $N=4$, the trajectory is tangent to the horizontal line $\ell=0$, and the two solutions for $\varphi_{\text{S}}^{2}$ in \eqref{BiON2} become degenerate. The plots are made at $k=1$ with $\e=0.3$.}
	\label{CF-plot-zoomin}
\end{figure}

Using \eqref{trajectory}, we plot the trajectories in the real ($\D-d/2$,$\,\ell$)-plane.
A simple example of Chew-Frautschi plot with $k=1, N=1$ can be found in Fig. \ref{CF-plot}.  
For generic $N$, the singlet trajectory has two different intersections with the horizontal line $\ell=0$. 
However, there exists a special value
\begin{align}
\label{Nast-leading}
N^\ast|_{\text{generic}\, k}=4+O(\e)\,,
\end{align}
at which the two intersections coincide and become degenerate at $\Delta=d/2$, 
as noticed above from \eqref{BiON2}.
To see the $\ell=0$ intersections more clearly, we zoom in on the region near $\ell=0$ for $k=1$ in Fig. \ref{CF-plot-zoomin}, 
where $N$ ranges from 2 to 6.
For generic $k$, we find:
\begin{itemize}
	\item
	After resolving the mixing of Regge trajectories near the leading intercept, 
	there exist two solutions to $\D=d/2$: 
	\begin{align}
	\ell_{\pm}^\ast=\(\frac 1 2 \pm \frac{\sqrt{6(N+2)}}{N+8}\)\,\e+O(\e^2)\,,
	\end{align}
	which is independent of $k$ at leading order. 
	The higher solution $\ell_{+}^\ast$ is known as the Regge intercept. 
	As we increase $N$, the distance between the two intercepts grows for $N<N^\ast$,  
	but decreases for $N>N^\ast$. 
	In particular, the lower intercept $\ell_{-}^\ast$ vanishes at the transition value $N=N^\ast$ 
	and thus the spin-0 intersections coincide. 
	\item
	As $N$ increases, the physical intersection with the $\ell=0$ line moves smoothly from left to right, 
	which is possible due to the vanishing lower intercept $\ell_-^\ast$ at the transition value $N=N^\ast$. 
	Accordingly, the scaling dimension of the singlet operator $\varphi_{\text{S}}^{2}$ is smaller than $d/2$ for $N<N^\ast$ and greater than $d/2$ for $N>N^\ast$. 
\end{itemize}
In general, we can define the transition value $N^\ast$ by the degeneracy condition 
$\D_{\varphi_{\text{S}}^{2}}=d/2$. 
In the canonical case $k=1$, 
we can make use of the high order expression of $\D_{\varphi_{\text{S}}^{2}}$ derived in the traditional diagrammatic approach \cite{Henriksson:2022rnm}. 
The corresponding $k=1$ transition value is
\begin{align}
\label{Nastk1}
N^\ast|_{k=1}=&\,4-4\e+\frac{1+7\zeta_3}{2}\e^2+\frac{31-89\zeta_3+189\zeta_4-730 \zeta_5}{72}\e^3
\nn&\,
+\frac{-140 + 709 \zeta_3 - 4824 \zeta_3^2 - 3204 \zeta_4 + 14298 \zeta_5- 43800 \zeta_6 + 
 110691 \zeta_7}{3456}\e^4+O(\e^5)\,,
\end{align}
where $\z_{i}=\z(i)$ denotes the value of the Riemann zeta function at integer $i$.
Some higher-order terms can also be computed easily, but they are not presented here for brevity. 
The transition value $N^\ast$ is not necessarily an integer due to the existence of higher order terms in $\e$. 
Using the order $\e^4$ expression for the singlet currents \cite{Manashov:2017xtt} , we can also derive the higher-order terms of the two intercepts at $k=1$:
\begin{align}
	\ell_{\pm}^\ast|_{k=1}=&\(\frac 1 2 \pm \frac{\sqrt{6(N+2)}}{N+8}\)\,\e
	-\(\frac{7(N+2)}{2(N+8)^2}\pm \frac{(5N^2+26N+68)}{(N+8)^3}\sqrt{\frac{N+2}{6}}\)\e^2
	\nn
	&+\Bigg(\frac{N+2}{8(N+8)^4}\Big(272 + 152 N + 41 N^2+16 (N + 8)^2\,\zeta_2\Big)
	\nn
	&\quad\pm
	\Scale[0.98]{\frac{1}{12(N+8)^5}\sqrt{\frac{N+2}{6}} }\Big(
	18752 + 14080 N + 1524 N^2 - 254 N^3 - N^4+12 (N+8)^4\,\zeta_2 
	\nn
	&\qquad\qquad\qquad\qquad\qquad\qquad\quad+72 (N-4) (N+8) (5N+22)\,\zeta_3\Big)
	\Bigg)\e^{3}+O(\e^4)\,,
\end{align}
which are consistent with \cite{Henriksson:2022rnm,Caron-Huot:2022eqs}. 
To determine the extreme value of the lower intercept, we impose that 
$\ell^\ast_{-}|_{k=1}$ is stationary with respect to $N$:
\begin{align}
\frac{\pa}{\pa N^{(i)}}\ell^\ast_{-}|_{k=1}=0,\quad i=0,1,2,\dots,
\end{align}
where $N^{(i)}$ is defined by the perturbative expansion $N=\sum_{i=0}^\infty N^{(i)}\,\e^i$. 
The solution $N^{(0)}=4, N^{(1)}=-4, N^{(2)}=(1+7\zeta_3)/2$ gives precisely the first three coefficients of  $N^{\ast}|_{k=1}$ in \eqref{Nastk1}. 
The corresponding extreme value 
\begin{align}
\ell_{-}^\ast|_{k=1,\text{extreme}}=O(\e^4)
\end{align}
is consistent with our general expectation that $\ell_-^\ast|_{N=N^\ast}=0$. 

\subsection{Analytic bootstrap}
\label{sec:Lorentzian Inversion2}
We again use the Lorentzian inversion formula to verify the consistency of the multiplet recombination results with OPE associativity. 
The four-point function of fundamental fields reads
\begin{align}
	\<\varphi_{a}(x_{1})\varphi_{b}(x_{2})\varphi_{c}(x_{3})\varphi_{e}(x_{4})\>=\frac{\mathcal{G}_{abce}(u,v)}{x_{12}^{2\Delta_{\varphi}}x_{34}^{2\Delta_{\varphi}}}\,.
\end{align}
As the intermediate operators transform in three irreducible representations of the O($N$) symmetry, 
we introduce the basis of tensor structures:
\begin{align}
	\mathcal{G}_{abce}(u,v)=\mathcal{G}_{\text{S}}(u,v)\mathbf{T}_{abce}^{\text{S}}+\mathcal{G}_{\text{T}}(u,v)\mathbf{T}_{abce}^{\text{T}}+\mathcal{G}_{\text{A}}(u,v)\mathbf{T}_{abce}^{\text{A}}\,,
\end{align}
where
\begin{align}
	\mathbf{T}_{abce}^{\text{S}}=\delta_{ab}\delta_{ce}\,,\qquad\mathbf{T}_{abce}^{\text{T}}=\frac{\delta_{ac}\delta_{be}+\delta_{ae}\delta_{bc}}{2}-\frac{1}{N}\delta_{ab}\delta_{ce}\,,\qquad\mathbf{T}_{abce}^{\text{A}}=\frac{\delta_{ac}\delta_{be}-\delta_{ae}\delta_{bc}}{2}\,.
\end{align}
The $1\leftrightarrow3$ crossing equation $v^{\Delta_{\varphi}}\mathcal{G}_{abce}(u,v)=u^{\Delta_{\varphi}}\mathcal{G}_{cbae}(v,u)$ results in three crossing equations for the three tensor structures respectively:
\begin{align}
	v^{\Delta_{\varphi}}\mathcal{G}_{\text{S}}(u,v)=&\;u^{\Delta_{\varphi}}\left(\frac{1}{N}\mathcal{G}_{\text{S}}(v,u)+\frac{(N-1)(N+2)}{2N^{2}}\mathcal{G}_{\text{T}}(v,u)-\frac{N-1}{2N}\mathcal{G}_{\text{A}}(v,u)\right)\,,\\
	v^{\Delta_{\varphi}}\mathcal{G}_{\text{T}}(u,v)=&\;u^{\Delta_{\varphi}}\left(\mathcal{G}_{\text{S}}(v,u)+\frac{N-2}{2N}\mathcal{G}_{\text{T}}(v,u)+\frac{1}{2}\mathcal{G}_{\text{A}}(v,u)\right)\,,\label{517}\\
	v^{\Delta_{\varphi}}\mathcal{G}_{\text{A}}(u,v)=&\;u^{\Delta_{\varphi}}\left(-\mathcal{G}_{\text{S}}(v,u)+\frac{N+2}{2N}\mathcal{G}_{\text{T}}(v,u)+\frac{1}{2}\mathcal{G}_{\text{A}}(v,u)\right)\,.\label{518}
\end{align}

To order $\e^{2}$, we need to consider the cross-channel scalars $\varphi_\text{S,T}^{2(n-1)}$ in the singlet and the symmetric traceless representations
\begin{align}
\varphi_\text{S}^{2(n-1)}=\varphi^{2(n-1)}\,,\qquad
\varphi_{\text{T},ab}^{2(n-1)}=\(\varphi_{a}\varphi_{b}-\frac{\d_{ab}}{N}\varphi^2\)\varphi^{2(n-2)}\,.
\end{align}
The anti-symmetric counterpart of $\varphi^{2q}$ does not exist.
We use the same classification of the two types of inversions as in the $N=1$ case: 
\begin{itemize}
\item
Type I: $n=2$

The double discontinuity at order $\epsilon^{2}$ 
involves the anomalous dimensions of $\varphi^{2}_{\text{S}}$ and $\varphi^{2}_{\text{T}}$
\begin{equation}
	\gamma_{\varphi^{2}_{\text{S}}}=\frac{N+2}{N+8}\epsilon+O(\epsilon^{2})\,,\qquad\gamma_{\varphi^{2}_{\text{T}}}=\frac{2}{N+8}\epsilon+O(\epsilon^{2})\,,
\end{equation}
where $\gamma_{\varphi^{2}_{\text{S}}}=\g_{2}^{\phantom{()}\!\!}$ has been derived in \eqref{ON-phi-2a} and the computation of $\g_{\varphi_{\text{T}}^{2}}$ can be found in footnote \ref{phi2-T}.
The inversion procedure is similar to the $N=1$ case.
The results are consistent with those from the multiplet recombination, and we have checked this for the cases of $m=0,1,2$.

\item
Type II: $n>2$ ($k$ and $n-1$ have no common divisor)

To order $\epsilon^{2}$, the operators contributing to the double discontinuity are $\varphi^{2n-2}_{\text{S}}$ and $\varphi^{2n-2}_{\text{T}}$, with OPE coefficients obtained in Appendix \ref{sec:b}:
\begin{align}	
\tilde \l^2_{\varphi\varphi\varphi^{2n-2}_{\text{S}}}=\frac{\l^{2}_{\varphi\varphi\varphi^{2n-2}_{\text{S}}}}{4^{n-1}(n-1)!\(\frac{N}{2}\)_{n-1}}\,,\quad
\tilde \l^2_{\varphi\varphi\varphi^{2n-2}_{\text{T}}}=\frac{\l^{2}_{\varphi\varphi\varphi^{2n-2}_{\text{T}}}}{2^{2n-3}(n-2)!\(\frac{N}{2}+2\)_{n-2}}\,.
\end{align}
The inversion results are also compatible with the multiplet recombination results,
which is verified for $m=0,1,2$.
\end{itemize}
Therefore, we confirm the consistency of the multiplet recombination and analytic bootstrap results for O($N$) models.

\section{Discussion}
In this work, we have studied the higher-derivative generalizations of the Ising multicritical CFTs, i.e., the $\Box^{k}$ scalar CFTs deformed by $\phi^{2n}$ interactions, and their O($N$) extensions. 
We derived the general formulas \eqref{gamma-spin-generic} and \eqref{gamma-spin-special} for the leading anomalous dimensions of conserved and partially conserved higher-spin currents. 
We first computed them from the multiplet recombination and then used the Lorentzian inversion formula to verify that they are compatible with crossing symmetry. In addition, we extended the results to the O($N$) models in \eqref{BiON1} and \eqref{BiON2}. 
In the O($N$) case, we further discussed the $N$ dependence of the Chew-Frautschi plot 
and pointed out the existence of a special value $N^\ast$ at which the two spin-0 intersections become degenerate. 
The explicit expressions of $N^\ast$ are given in \eqref{Nast-leading} and \eqref{Nastk1}. 
We plan to study the higher-derivative defect CFTs and deduce the defect CFT data for generic $\{k, n, N\}$. 

Besides the spinless multiplet recombination \eqref{rec}, there exist spinning multiplet recombination phenomena. The shortening condition for the spinning current $\mathcal{J}_{\ell}^{(m)}$ is broken at the generalized WF fixed points, 
so $\mathcal{J}_{\ell}^{(m)}$ should recombine with another spinning multiplet to form a long multiplet, 
except for the stress tensor and the O($N$) global symmetry currents. 
The $k=1$ case had been studied in \cite{Skvortsov:2015pea,Giombi:2016hkj,Roumpedakis:2016qcg,Giombi:2017rhm}.
It is curious to see how the multiplets of partially conserved currents recombine 
in the generalized WF CFTs.

It would be interesting to investigate these higher-derivative multicritical CFTs at integer dimensions if they indeed exist. 
As the Gaussian CFTs with higher derivatives are nonunitary, 
we expect that the interacting CFTs violate unitarity as well. 
To perform the nonperturbative study, 
we need to use the conformal bootstrap methods that do not rely on positivity constraints 
\cite{Gliozzi:2013ysa,Gliozzi:2014jsa, El-Showk:2016mxr, Esterlis:2016psv, Li:2017ukc,Li:2021uki, Kantor:2021kbx, Kantor:2021jpz, Afkhami-Jeddi:2021iuw,Laio:2022ayq}.

\section*{Acknowledgments}
We would like to thank J. Henriksson for helpful correspondence. We also thank the referee for the valuable comments and constructive suggestions. 
This work was supported by the 100 Talents Program of Sun Yat-sen University and the
Natural Science Foundation of China (Grant No. 12205386) and the Guangzhou Municipal
Science and Technology Project (Grant No. 2023A04J0006).

\appendix

\section{Some OPE coefficients}
\label{sec:b}
The three-point function coefficients $\l_{\phi\phi\phi^{2(n-1)}},\l_{\phi\phi\phi^{2n}}$ are constrained by the matching conditions \eqref{match-3}. 
The solutions are
\begin{align}
	\l_{\phi\phi\phi^{2(n-1)}}&=-\frac{k!(n!)^{3}\left(1+\frac{k}{n-1}\right)_{k-1}}{4^{k-1}(2n)!(\frac{3}{2})_{k-1}\left(-\frac{(n-2)k}{n-1}\right)_{k}}\,\epsilon+O(\epsilon^{2})\,,
	\\
	\l_{\phi\phi\phi^{2n}}&=-\frac{(n!)^{3}\left(1+\frac{k}{n-1}\right)_{k-1}}{(2n-1)!\left(1-\frac{(2n-1)k}{n-1}\right)_{k}}\,\epsilon+O(\epsilon^{2})\,.
\end{align}
The type II inversion in O($N$) models involves the OPE coefficients $\l_{\varphi_{a}\varphi_{b}\varphi^{2n-2}_{\text{S}}}$ and $\l_{\varphi_{a}\varphi_{b}\varphi^{2n-2}_{\text{T}}}$.
We consider the O($N$) generalizations of \eqref{boxk-1-1-2n}
\begin{align}
	\lim_{\a\rightarrow0}\alpha_{N}^{-1}\<\Box^{k}\varphi_{a}(x_{1})\varphi_{b}(x_{2})\varphi^{2n-2}_{\text{S}}(x_{3})\>&=\<\varphi_{a}\varphi^{2n-2}(x_{1})\varphi_{b}(x_{2})\varphi^{2n-2}_{\text{S}}(x_{3})\>_{\text{f}}\,,\\
	\lim_{\a\rightarrow0}\alpha_{N}^{-1}\<\Box^{k}\varphi_{a}(x_{1})\varphi_{b}(x_{2})\varphi^{2n-2}_{\text{T},ce}(x_{3})\>&=\<\varphi_{a}\varphi^{2n-2}(x_{1})\varphi_{b}(x_{2})\varphi^{2n-2}_{\text{T},ce}(x_{3})\>_{\text{f}}\,.
\end{align}
In the basis $\d_{ab}$ and $\frac{\delta_{ac}\delta_{be}+\delta_{ae}\delta_{bc}}{2}-\frac{1}{N}\delta_{ab}\delta_{ce}$ for the singlet and symmetric traceless cases respectively, we obtain
\begin{align}
	\l_{\varphi\varphi\varphi^{2n-2}_{\text{S}}}&=\frac{4^{n-2k}(n-1)!\(1+\frac{N}{2}\)_{n-1}}{\(\frac{3}{2}\)_{k-1}\(1-\frac{k}{n-1}\)_{k}}\a_{N}+O(\e^{2})\,,\\
	\l_{\varphi\varphi\varphi^{2n-2}_{\text{T}}}&=\frac{4^{n-2k}(n-1)!\(2+\frac{N}{2}\)_{n-2}}{\(\frac{3}{2}\)_{k-1}\(1-\frac{k}{n-1}\)_{k}}\a_{N}+O(\e^{2})\,.
\end{align}

\section{Ratios of three-point function coefficients}
\label{sec:Ratio of 3 point function coefficients}
The ratio of three-point function coefficients \eqref{ratio-3pt-current} is obtained by considering the following four-point functions
\begin{align}
	\<\phi_{\text{f}}(x_{1})\phi_{\text{f}}(x_{2})\phi_{\text{f}}(x_{3})\phi_{\text{f}}(x_{4})\>&=\frac{\mathcal{G}_{1}(u,v)}{x_{12}^{2\D_{\phi_\text{f}}} x_{34}^{2\D_{\phi_{\text{f}}}}}\,,
	\\
	\<\phi^{2n-1}_{\text{f}}(x_{1})\phi^{2n-1}_{\text{f}}(x_{2})\phi_{\text{f}}(x_{3})\phi_{\text{f}}(x_{4})\>&=\frac{\mathcal{G}_{2}(u,v)}{x_{12}^{2(2n-1)\D_{\phi_\text{f}}} x_{34}^{2\D_{\phi_{\text{f}}}}}\,.
\end{align}
where $u=z\bar{z}$ and $v=(1-z)(1-\bar{z})$.
Recall that we have normalized the correlators such that the three-point function coefficients are exactly given by Wick contractions
\begin{align}
	\mathcal{G}_{1}(u,v)&=1+u^{\Delta_{\phi_{\text{f}}}}\left(1+\frac{1}{v^{\Delta_{\phi_{\text{f}}}}}\right)\,,\\
	\mathcal{G}_{2}(u,v)&=(2n-1)!+(2n-1)^{2}(2n-2)!u^{\Delta_{\phi_{\text{f}}}}\left(1+\frac{1}{v^{\Delta_{\phi_{\text{f}}}}}\right)\,,
\end{align}
which have the same $u,v$ dependencies up to the factor $(2n-1)^{2}(2n-2)!$. When expanded into conformal blocks, the three-point function coefficients should satisfy
\begin{align}
	\l_{\phi^{2n-1}_{\text{f}}\phi^{2n-1}_{\text{f}}\mathcal{J}_{\ell,\text{f}}^{(m)}}\l_{\phi_{\text{f}}^{}\,\phi_{\text{f}}^{}\,\mathcal{J}_{\ell,\text{f}}^{(m)}}=(2n-1)^{2}(2n-2)!\l_{\phi_{\text{f}}^{}\,\phi_{\text{f}}^{}\,\mathcal{J}_{\ell,\text{f}}^{(m)}}^{2}\,,
\end{align}
which gives the ratios of the three-point function coefficients \eqref{ratio-3pt-current}.

For the O($N$) model, we consider four-point functions
\begin{align}
	\<\varphi_{a}(x_{1})\varphi_{b}(x_{2})\varphi_{c}(x_{3})\varphi_{e}(x_{4})\>_{\text{f}}&=\frac{\mathcal{G}_{1}(u,v)}{x_{12}^{2\D_{\varphi_\text{f}}} x_{34}^{2\D_{\varphi_{\text{f}}}}}\,,
	\\
	\<\varphi_{a}\varphi^{2n-2}(x_{1})\varphi_{b}\varphi^{2n-2}(x_{2})\varphi_{c}(x_{3})\varphi_{e}(x_{4})\>_{\text{f}}&=\frac{\mathcal{G}_{2}(u,v)}{x_{12}^{2(2n-1)\D_{\varphi_\text{f}}} x_{34}^{2\D_{\varphi_{\text{f}}}}}\,.
\end{align}
We introduce the following basis of tensor structures:
\begin{align}
	\mathcal{G}(u,v)=\mathcal{G}_{\text{S}}(u,v)\mathbf{T}_{abce}^{\text{S}}+\mathcal{G}_{\text{T}}(u,v)\mathbf{T}_{abce}^{\text{T}}+\mathcal{G}_{\text{A}}(u,v)\mathbf{T}_{abce}^{\text{A}}\,,
\end{align}
where
\begin{align}
	\mathbf{T}_{abce}^{\text{S}}=\delta_{ab}\delta_{ce}\,,\qquad\mathbf{T}_{abce}^{\text{T}}=\frac{\delta_{ac}\delta_{be}+\delta_{ae}\delta_{bc}}{2}-\frac{1}{N}\delta_{ab}\delta_{ce}\,,\qquad\mathbf{T}_{abce}^{\text{A}}=\frac{\delta_{ac}\delta_{be}-\delta_{ae}\delta_{bc}}{2}\,.
\end{align}
So the $\mathcal{G}$ functions are
\begin{align}
	\mathcal{G}_{\text{S},1}(u,v)&=1+\frac{1}{N}u^{\D_{\varphi_{\text{f}}}}\left(1+\frac{1}{v^{\D_{\varphi_{\text{f}}}}}\right)\,,\\
	\mathcal{G}_{\text{S},2}(u,v)&=4^{n-1}(n-1)!\left(\frac{N}{2}+1\right)_{n-1}
	\nn
	&\phantom{=}+\frac{4^{n-1}(2n-1)(n-1)!\left(\frac{N}{2}+1\right)_{n-1}}{N}u^{\D_{\varphi_{\text{f}}}}\left(1+\frac{1}{v^{\D_{\varphi_{\text{f}}}}}\right),\\
	\mathcal{G}_{\text{T},1}(u,v)&=u^{\D_{\varphi_{\text{f}}}}\left(1+\frac{1}{v^{\D_{\varphi_{\text{f}}}}}\right)\,,\\
	\mathcal{G}_{\text{T},2}(u,v)&=2^{2n-3}(4n+N-2)(n-1)!\left(\frac{N}{2}+2\right)_{n-2}u^{\D_{\varphi_{\text{f}}}}\left(1+\frac{1}{v^{\D_{\varphi_{\text{f}}}}}\right)\,,\\
	\mathcal{G}_{\text{A},1}(u,v)&=u^{\D_{\varphi_{\text{f}}}}\left(1-\frac{1}{v^{\D_{\varphi_{\text{f}}}}}\right)\,,\\
	\mathcal{G}_{\text{A},2}(u,v)&=4^{n-1}(n-1)!\left(\frac{N}{2}+1\right)_{n-1}u^{\D_{\varphi_{\text{f}}}}\left(1-\frac{1}{v^{\D_{\varphi_{\text{f}}}}}\right)\,,
\end{align}
where the $u,v$ dependencies of the two correlators are the same up to a factor for each component. Then we can derive the ratios of three-point function coefficients in \eqref{RaBiON}.

\section{Light cone expansion of conformal blocks}
\label{Lightcone expansion of conformal blocks}
Consider the four-point function $\langle \phi(x_{1})\phi(x_{2})\phi(x_{3})\phi(x_{4}) \rangle$.
The conformal blocks satisfy the Casimir equation \cite{Dolan:2003hv}
\begin{equation}
	\mathcal{C}_{2}G^{(d)}_{\Delta,\ell}(z,\bar{z})=\left(h(h+1-d)+\bar{h}(\bar{h}-1)\right)G^{(d)}_{\Delta,\ell}(z,\bar{z}),
\end{equation}
where the quadratic Casimir operator is
\begin{equation}
	\mathcal{C}_{2}=D_{z}+D_{\bar{z}}+(d-2)\frac{z\bar{z}}{z-\bar{z}}\left((1-z)\partial_{z}-(1-\bar{z})\partial_{\bar{z}}\right),
\end{equation}
with
\begin{equation}
	D_{z}=z^{2}\partial_{z}(1-z)\partial_{z}.
\end{equation}
The boundary condition is $G^{(d)}_{\Delta,\ell}(z,\bar{z})=z^{h}\bar{z}^{\bar{h}}(1+\ldots)$ as $z\rightarrow0$, $\bar{z}\rightarrow0$. 
The small $z$ expansion of conformal block reads \cite{Simmons-Duffin:2016wlq}:
\begin{equation}\label{424}
	G^{(d)}_{\Delta,\ell}(z,\bar{z})=\sum_{i=0}^{\infty}z^{h+i}\sum_{j=-i}^{i}c_{i,j}^{(d)}(h,\bar{h})\bar{z}^{\bar{h}+j}\,_{2}F_{1}(\bar{h}+j,\bar{h}+j,2(\bar{h}+j),\bar{z})\,,
\end{equation}
where $h=\frac{\Delta-\ell}{2}$, $\bar{h}=\frac{\Delta+\ell}{2}$, and $_{2}F_{1}$ is the Gaussian hypergeometric function.  
The coefficient $c_{i,j}^{(d)}(h,\bar{h})$ can be determined by solving the Casimir equation order by order in $z$, 
To study the leading, subleading and sub-subleading trajectory, we need to know the coefficient $c_{i,j}^{(d)}(h,\bar{h})$ with $i=0,1,2$. 
Their explicit expressions are
\begin{align}
	\label{normalization}
	c_{0,0}^{(d)}(h,\bar{h})=\;&1,\\
	c_{1,-1}^{(d)}(h,\bar{h})=\;&-\frac{(d-2)(h-\bar{h})}{d-2h+2\bar{h}-4},\\
	c_{1,0}^{(d)}(h,\bar{h})=\;&\frac{h}{2},\\
	c_{1,1}^{(d)}(h,\bar{h})=\;&\frac{(d-2)\bar{h}^{2}(h+\bar{h}-1)}{4(2\bar{h}-1)(2\bar{h}+1)(2-d+2h+2\bar{h})},\\
	c_{2,-2}^{(d)}(h,\bar{h})=\;&\frac{(d-2)d(h-\bar{h})(h-\bar{h}+1)}{2(d-2h+2\bar{h}-6)(d-2h+2\bar{h}-4)},\\
	c_{2,-1}^{(d)}(h,\bar{h})=\;&-\frac{(d-2)(h+1)(h-\bar{h})}{2(d-2h+2\bar{h}-4)},\\
	c_{2,0}^{(d)}(h,\bar{h})=\;&-\frac{1}{32}(d-2h-3)(d+2h)+\frac{1}{64}(3d^{2}-10d+6)\nn
	&-\frac{(2\bar{h}-3)(2\bar{h}+1)}{32(d-2h-3)(d-2h-2\bar{h}-2)(d-2h+2\bar{h}-4)}\nn
	&+\frac{d^{2}-6d+6}{16(d-2h-3)(d-2h-2\bar{h}-2)(d-2h+2\bar{h}-4)}\nn
	&+\frac{(d-2)d(d-2h-5)(2h-3)(2h-1)}{64(d-2h-3)(d-2h-2\bar{h}-2)(2\bar{h}-3)(2\bar{h}+1)(d-2h+2\bar{h}-4)}\nn
	&+\frac{(-d+2h+5)(d^{3}-8d^{2}+20d-17)}{32(d-2h-2\bar{h}-2)(d-2h+2\bar{h}-4)}\nn
	&+\frac{d^{4}-16d^{3}+83d^{2}-174d+132}{64(d-2h-2\bar{h}-2)(d-2h+2\bar{h}-4)},\\
	c_{2,1}^{(d)}(h,\bar{h})=\;&-\frac{(d-2)(h+1)\bar{h}^{2}(h+\bar{h}-1)}{8(d-2h-2\bar{h}-2)(2\bar{h}-1)(2\bar{h}+1)},\\
	c_{2,2}^{(d)}(h,\bar{h})=\;&\frac{(d-2)d\bar{h}^{2}(\bar{h}+1)^{2}(h+\bar{h}-1)(h+\bar{h})}{32(d-2h-2\bar{h}-4)(d-2h-2\bar{h}-2)(2\bar{h}-1)(2\bar{h}+1)^{2}(2\bar{h}+3)}\,.
\end{align}

\section{Inversion at subleading and sub-subleading twists}
\label{Subleading twist and sub-subleading twist}
In this appendix, we discuss the inversion procedure for the subleading and sub-subleading trajectories.
We consider the scalar block \cite{Li:2020ijq}
\begin{align}\label{scalar-block-z2}
	&G^{(d)}_{\D,0}(1-\bar{z},1-z)\big|_{\log z}\nn
	=&\;(1-\bar{z})^{\D/2}\frac{\Gamma(\D)}{\Gamma(\D/2)^{2}}\Big\{-\,_{2}F_{1}\(\frac{\D}{2},\frac{\D}{2},1-\frac{d}{2}+\D,1-\bar{z}\)\nn
	&+\frac{1}{4}\D\[(2-\D)\,_{2}F_{1}\(\frac{\D}{2},\frac{\D}{2},1-\frac{d}{2}+\D,1-\bar{z}\)\right.\nn
	&+\left.\frac{(d-2)\D}{d-2(\D+1)}(1-\zb)\,_{2}F_{1}\(1+\frac{\D}{2},1+\frac{\D}{2},2-\frac{d}{2}+\D,1-\bar{z}\)\]z\nn
	&+\frac{1}{64}\D\[(\D-2)(-\D^{2}+2\D-8)\,_{2}F_{1}\(\frac{\D}{2},\frac{\D}{2},1-\frac{d}{2}+\D,1-\bar{z}\)\right.\nn
	&+\frac{2\D(\D^{2}+4)(d-2)}{d-2(\D+1)}(1-\zb)\,_{2}F_{1}\(1+\frac{\D}{2},1+\frac{\D}{2},2-\frac{d}{2}+\D,1-\bar{z}\)\nn
	&\left.-\frac{d(d-2)\D(\D+2)^{2}}{(d-2(\D+1))(d-2(\D+2))}(1-\zb)^{2}\,_{2}F_{1}\(2+\frac{\D}{2},2+\frac{\D}{2},3-\frac{d}{2}+\D,1-\bar{z}\)\]z^{2}\nn
	&+O(z^{3})\Big\}\,,
\end{align}
and the crossing factor
\begin{align}\label{crosing-factor-z2}
	\frac{z^{\D_{\phi}}\zb^{\D_{\phi}}}{ (1-z)^{\D_{\phi}}(1-\zb)^{\D_{\phi}} }\,.
\end{align}
As in the leading trajectory calculation, there are also type I and II inversions in the subleading and sub-subleading cases.
We substitute $\D=2\D_{\phi}+\tilde{\g}_{\phi^{2}}$ into \eqref{scalar-block-z2} for the type I inversion, and we substitute $\D=2k+O(\e)$ into \eqref{scalar-block-z2} for the type II inversion.
After multiplying the results by the crossing factor \eqref{crosing-factor-z2}, we evaluate the double discontinuities.

In the cases of higher trajectories, the anomalous dimensions are given by the inversion integral \eqref{SL2R-inversion} at higher orders in $z$.
For subleading twist calculations, we are interested in the inversion at order $z^{\D_{\phi}+1}$.
The $\log z$ part of the inversion integral in \eqref{SL2R-inversion} corresponds to
\begin{align}
	\frac{1}{2}c_{0,0}(h+1,\bar{h})\(\tilde{\l}^2_{\phi\phi\mathcal{J}_{\ell}^{(1)}}\tilde{\gamma}_{\mathcal{J}_{\ell}^{(1)}}\)_{\ell\rightarrow\bar{h}-\D_{\phi}}+\frac{1}{2}\sum_{j=-1}^{1}c_{1,j}(h,\bar{h}-j)\(\tilde{\l}^2_{\phi\phi\mathcal{J}_{\ell}^{(0)}}\tilde{\gamma}_{\mathcal{J}_{\ell}^{(0)}}\)_{\ell\rightarrow\bar{h}-j-\D_{\phi}}\,.
\end{align}
The result contains the contributions from the leading trajectory.
We obtain the anomalous dimensions $\tilde{\gamma}_{\mathcal{J}_{\ell}^{(1)}}$ of bilinear operators in the subleading twist trajectory after subtracting the leading trajectory contribution.
The sub-subleading twist calculations are similar.
We are then interested in the inversion at order $z^{\D_{\phi}+2}$.
The result of the inversion integral corresponds to the terms
\begin{align}
	&\frac{1}{2}c_{0,0}(h+2,\bar{h})\(\tilde{\l}^2_{\phi\phi\mathcal{J}_{\ell}^{(2)}}\tilde{\gamma}_{\mathcal{J}_{\ell}^{(2)}}\)_{\ell\rightarrow\bar{h}-\D_{\phi}}\nn
	&+\frac{1}{2}\sum_{j_{2}=-1}^{1}c_{1,j_{2}}(h+1,\bar{h}-j_{2})\(\tilde{\l}^2_{\phi\phi\mathcal{J}_{\ell}^{(1)}}\tilde{\gamma}_{\mathcal{J}_{\ell}^{(1)}}\)_{\ell\rightarrow\bar{h}-j_2-\D_{\phi}}\nn
	&+\frac{1}{2}\sum_{j_{1}=-2}^{2}c_{2,j_{1}}(h,\bar{h}-j_{1})\(\tilde{\l}^2_{\phi\phi\mathcal{J}_{\ell}^{(0)}}\tilde{\gamma}_{\mathcal{J}_{\ell}^{(0)}}\)_{\ell\rightarrow\bar{h}-j_{1}-\D_{\phi}}\,,
\end{align}
and we obtain the anomalous dimensions $\tilde{\gamma}_{\mathcal{J}_{\ell}^{(2)}}$ of bilinear operators in the sub-subleading twist trajectory.
All the results agree with those from the multiplet recombination method.

\end{document}